# Talent Recommendation on LinkedIn User Profiles

**Yuzhou Peng**

## A Dissertation

Presented to the University of Dublin, Trinity College

in partial fulfilment of the requirements for the degree of

## Master of Science in Computer Science (Intelligent System)

Supervisor: Professor Seamus Lawless,

August 2018

# Declaration

I, the undersigned, declare that this work has not previously been submitted as an exercise for a degree at this, or any other University, and that unless otherwise stated, is my own work.

_________________________________________

Yuzhou Peng

August 30, 2018

# Permission to Lend and/or Copy

I, the undersigned, agree that Trinity College Library may lend or copy this thesis upon request.

_________________________________________

Yuzhou Peng

August 30, 2018

# Acknowledgments

I would first like to thank Dr. Seamus Lawless and Yu Xu for their kind help in my dissertation.

I would also like to thank lecturers and staffs from School of Computer Science and Statistics and class fellow for their helps in my postgradulate study.

Finally, I would like to express my gratitude to my family sincerely because I could not start my further study without their support.

Yuzhou Peng

*University of Dublin, Trinity College*
*August 2018*

# Talent Recommendation on LinkedIn User Profiles

Yuzhou Peng , Master of Science in Computer Science

University of Dublin, Trinity College, 2018

Supervisor: Professor Seamus Lawless,

With the increasing amount of information on the Internet, recommender systems are becoming increasingly crucial in supporting people to find and explore relevant content. This is also true in the online recruitment space, with websites such as LinkedIn, Indeed.com and Monster.com all using recommender systems. In online recruitment, it can often be challenging for companies to find suitable candidates with appropriate skills because of the huge volume of users profiles available. Identifying users which satisfy a range of different employer needs is also a difficult task. Thus, effective matching of user profiles and jobs is becoming crucial for companies. This research project applies a wide range of recommendation techniques to the task of user profile recommendation. Extensive experiments are conducted on a large-scale real-world LinkedIn dataset to evaluate their performance, with the aim of identifying the most suitable approach in this particular recommendation scenario.

# Summary

The aim of this research project is to identify appropriate recommendation approaches for the task of user profile recommendation on online recruitment environments based on existing recommendation techniques. In particular, the research conducted on this project focuses on the popular professional recruitment platform: LinkedIn. On LinkedIn, the primary approach for users to find jobs is by making applications for positions using their resumes/profiles provided on the platform. Companies, on the other hand, often use LinkedIn's Talent Solutions, which aims to help find and engage the most appropriate candidates, and user recommendation is one of its main features. This research project first conducts extensive analysis on an large number of user profiles, i.e. resumes of LinkedIn users, and attempts to understand the distributions of needs of both job seekers and recruiters. Then, experiments are conducted on a large-scale LinkedIn dataset to evaluate the performance of a range of existing recommendation approaches. Detailed analysis of experimental results is then presented in order to identify the most suitable recommendation approach in this particular application scenario.

# Contents

# List of Tables

# List of Figures

# Chapter 1

# Introduction

This chapter will introduce and discuss the challenge of talent recommendation, specifically using the LinkedIn user profiles. We will first outline the motivation for this research, then we will detail the research question addressed by this project. Finally, the research objectives of the recommender system to be developed will be provided.

## 1.1   Motivation

Recommender systems have been widely used in many application scenarios on the modern web because of the booming information stream that must be navigated. This research focuses on the domain of recruitment, specifically Job and Candidate recommendation. From the company's perspective, an effective recommender system could greatly reduce the workload involved in finding and engaging with appropriate candidates from the large number of user profiles available on job websites, such as Indeed, Monster, and job.com. In recruitment websites, a recommender system acts like a headhunter, identifying information such as: which user is the most suitable candidate for a specific job? Which candidates skills best fit a particular company's needs? Which company is a good target for a certain user? From the company's perspective, they need employees with certain specialized skills and work experience, who best fit their advertised positions. Meanwhile, users of online job websites also hope to find their dream job, which could upgrade their skills, expand their career opportunities and allow them to achieve their goals.

Online talent recommendation has recently become an important research topic. There is a significant number of existing works in this area, such as finding talent through recommendation in social networks by text-matching (Balog et al.2006). However, these approaches tend to have focused on recommending talent from general information, instead of recommending jobs/candidates using professional user profiles/resumes that include information such as user work experience, education, and skills. Thus our research aims to examine the performance of existing recommendation techniques when applied to this novel challenge.

Thus our motivation is to improve the process of talent recommendation in online recruitment sites through the adaption and application of existing approaches from the recommender systems and machine learning literature.

## 1.2   Research Question

We will try to investigate effective approaches for recommending talent on LinkedIn to companies that are looking to hire for a specific job using LinkedIn user profiles. Thus, our research question is:

*To what extent can recommendation techniques support talent recommendation, based on LinkedIn user profiles?*

## 1.3   Project Objectives

This aim of this research project is to identify suitable recommendation approaches for the job/candidate recommendation task using professional user profiles/resumes, with a focus on the LinkedIn platform. In order to achieve this goal, this project has the following three major objectives:

1) *Experimental Dataset Construction and Analysis*. In this project, a large number of LinkedIn user profiles were harvested and used to construct an experimental dataset, in order to simulate the application scenario of job/candidate recommendation using user profiles in an online recruitment platform. On LinkedIn, a user profile consists of multiple fields, each of which details a particular aspect of information about the user, such as work experiences, educational experiences and skills, associations, etc. Thus, the specific aim of this project is to recommend suitable users for a particular position by using the structured textual information provided in their LinkedIn pro- files. In the experiments, we assume that the users who actually specified the first job experience in a job position are taken as ground truth users for the recommendation of that job position. Furthermore, an extensive analysis on the dataset is also needed to understand the needs of the proposed recommendation task, such as the distribution of job positions taken by the users and the distribution of the number of work/educa- tional experiences of users, and help to select and design appropriate recommendation approaches.

2) *Recommendation Approach Selection and Design*. Data Analysis on the constructed dataset illustrated the diversity of domains of job positions, e.g. there are over 200,000 different job positions discovered from the work experiences of the 158096 user profiles in the dataset, and the uncertainty in terms of the richness and completeness of provided user profiles, e.g. the number of work experiences a user specified in his/her profile ranges from 0 up to 7. Thus, it suggests that a learning-based recommendation approach would be well suited to the targeted application scenario, rather than rule-based or process-oriented approaches. This is because learning-based approaches have the advantage of automatically learning recommendation models for different domains of job positions, rather than separately designing different models for hundreds of thousands of job positions. Specifically, this project targets two categories of learning-based recommendation approaches in the proposed talent recommendation task: content-based recommendation approaches and collaborative filtering-based recommendation approaches. A content-based recommendation approach generally consists of two components: feature engineering based on the input content and selection of learning algorithms. A collaborative filtering-based approach also usually consists of two key components: construction of a rating matrix based on the input content and the selection of collaborative filtering algorithms. Thus, different feature engineering approaches, learning algorithms and collaborative filtering recommendation approaches need to be experimented with to examine the performance of the two categories of recommendation approaches.

3) *Experimental results and analysis*. Detailed experimental methods/steps need to be designed and implemented in order to identify the most suitable recommendation approach in the proposed talent recommendation task. They include defining the experimental environment, preprocessing of the dataset, selecting the evaluation metrics, and targeted evaluation aspects, e.g. the impact of the training size on the recommendation approaches, or the impact of the dimension of feature representation.

## 1.4 Project Outline

In the chapter2 we will present the literature review of the relevant researches about techniques that could be used in the job recommendation. We will discuss the state- of-the-art and limitations of those techniques.

Chapter 3 will present and discuss the type of machine learning, approaches for recommender system, the statistic LinkedIn user data used by this research, detail of the design of the recommender system approaches and description of techniques used.

In the chapter 4 we will present results generated by various approaches from chapter 3. Then results will be evaluated based on selected metrics. Finally, we will indicate which approach can support the candidate recommendation for jobs.

Chapter 5 describe the conclusion drawn from this research, we will conclude the performance of selected approaches and give the conclusion of the recommender system.

We will also discuss the future works of our project.

.

# Chapter 2

# Related works

## 2.1 Introduction

Talent is considered as a candidate's capacity to improve their future career (Ashton & Morton2005). We will focus on techniques used for recommender system because the aim of the research is mainly concerns identifying approaches suitable for candidate recommendation for jobs. The term recommender system was defined for a long time ago by the research of (Resnick & Varian1997). However, the term recommender system represent a broad research field and there is ambiguity over what knowledge and techniques are involved in the field of job recommendation. (Paparrizos et al.2011), stated the term job recommendation refers to recommending candidate for jobs with machine learning techniques. Thus, we will give an overview of current techniques suitable for the realm of job recommendation. Based on a literature review, we discover that many techniques in job/candidate recommendation have focused on machine learning and feature representation. These researches will be presented in this chapter, from techniques of experimental data construction to techniques of learning algorithms for recommendation. This chapter will conclude with a discussion of what we have learned for the design of our recommendation system.

## 2.2   Talent Recommendation Techniques in Researches of Recommender System

There are a variety of traditional methods for talent recommendation on the web. (Becerra-Fernandez 2006). surveyed a large amount of talent-locator systems on the web, and a number of talent locator approaches that have been proposed are similar to CONNEX, which is a database containing a huge amount of user profiles with which companies can find users with desirable skills or other talents through a web browser interface. However, these approaches lack context information extraction and they can not predict the future jobs of candidates. Thus data mining techniques for user data is crucial for our research project. Based on the boom in research into feature representation and machine learning, we can find effective tools for feature extraction. We have found many types of research about techniques that could be used for talent recommendation.

### 2.2.1   Techniques Used for Feature Representation

**N-grams for extracting context information of data**

Comparing with early data mining technique, N-grams could extract the context information. The term N-grams has coined by (Brown et al.1992). Then many researchers have implemented the n-gram for talent recommendation. In an early research of (Su et al.2000), the researchers have developed a recommender system for predicting requests of the possible web page that users want to browse. The recommender system predict request by analyzing raw and filtered user log file with the n-gram model. They used unigram to 4-gram. As a result, the N-grams has shown great improvements of precision of prediction users viewing request.2.1

Other researches, like (Ashok et al.2009), the researchers developed Bug advisor which is a recommender system for debugging in the large software development projects. The researcher designed a baseline approach based on a full-text search i.e. search full text of the error in the error report. The researcher analyzed the error report of the bug and extracted the context information using bag of words and call it the Bug feature search. The bug advisor improved 33% recall than the full-text search (e.g. find 33% more bugs compared with full-text set).

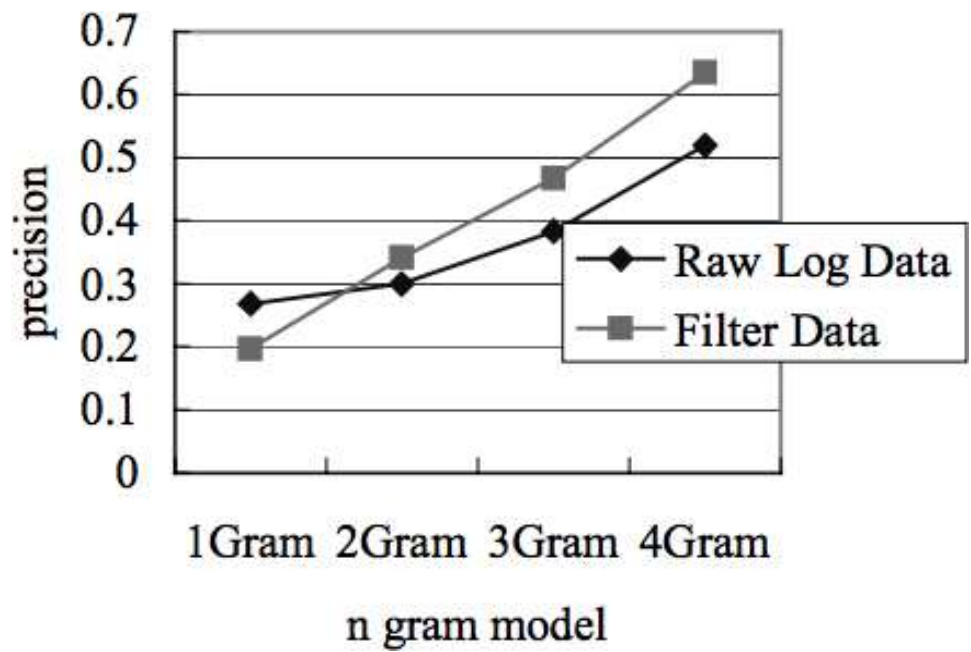

Figure 2.1: Comparing precision of using raw user log data and filtered user log data of different n-grams

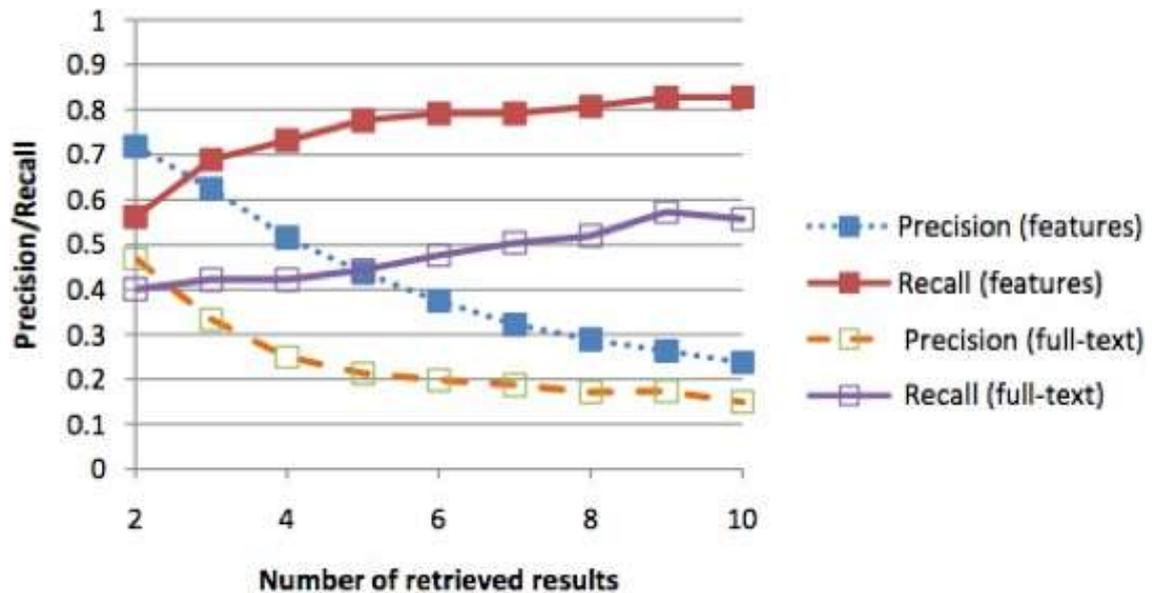

Figure 2.2: Precision and recall for two different approaches of Bug advisor

<span style="color:red">2.2</span>

In the research of (<span style="color:green">Stamatatos2011</span>), the researcher implemented the n-gram model for the detection of plagiarisms based on the context information of the user profiles. and the results showed that using n-grams could provide high accuracy results (0.79) of plagiarisms of the document.

**Word embedding for the expert recommendation**

In the above section, we have considered extract context information in the user profile using N-grams. More techniques for feature representation need to be proposed. Word embedding could provide word vectors that keeping semantic information inside the document. Thus, we will consider using word vectors to extract more semantic information of user profiles.

The research of (<span style="color:green">Mikolov et al.2013</span>) illustrates the process of the word embedding. Based on their theory, the word can be represented by word vectors by training huge corpus.

Word embedding have a number of implementation, the (<span style="color:green">Zheng et al.2017</span>) have conducted the word embedding for user rating prediction using DeepCoCNN neural network with collaborative filtering learning algorithms. The researcher use matrix factorization learning algorithms as the baseline. As a result, the MSE of the word embedding approach decreased more than 7% than baseline approaches.2.2.1

Table 3: **MSE Comparison with baselines. Best results are indicated in bold.**

| Dataset | MF | PMF | LDA | CTR | HFT-10 | HFT-50 | CDL | DeepCoNN | Improvement of DeepCoNN (%) |
|---|---|---|---|---|---|---|---|---|---|
| Yelp | 1.792 | 1.783 | 1.788 | 1.612 | 1.583 | 1.587 | 1.574 | **1.441** | 8.5% |
| Amazon | 1.471 | 1.460 | 1.459 | 1.418 | 1.378 | 1.383 | 1.372 | **1.268** | 7.6% |
| Beer | 0.612 | 0.527 | 0.306 | 0.305 | 0.303 | 0.302 | 0.299 | **0.273** | 8.7% |
| **Average on all datasets** | 1.292 | 1.256 | 1.184 | 1.112 | 1.088 | 1.09 | 1.081 | **0.994** | 8.3% |

### Document embedding for the expert recommendation

Document embedding could contain the semantic information of multiple documents with generated document vectors. The derivative of word embedding is the document embedding, document embedding introduce document vector to contain semantic information of words in different documents. In (Le & Mikolov2014), the research has proposed how to form document vectors based on doc2vec, a powerful tool for document vector construction derived from word2vec.

There are some implementations of doc2vec (Vasile et al.2016) has developed a system for author profiling, they use doc2vec for feature representation and logistic regression classifier. The researcher selected data of blog post and social media, then the researcher set the baseline of the research of (Modaresi et al.2016), bag of words with logistic regression classifier, bag of c3grams with logistic regression classifier. The researcher then tested the doc2vec and collaborative filtering approach. Finally, the researcher calculated the accuracy of approaches and compare doc2vec with baselines. The result shown that doc2vec performs best with accuracy of 42.20 compared with baselines in the age data of the blog post.2.3

### Dimension reduction techniques for feature representation

In order to lower the time costing and improve the usability of the recommender engine. We decided to implement dimension reduction techniques (e.g. Singular Value Decomposing) in our recommender system because the learning algorithms could be sensitive to high dimensional feature data.

| Approach | Blog posts | | Social media | |
|---|---|---|---|---|
| | Age | Gender | Age | Gender |
| Modaresi *et al.* [31] | 40.91 | **77.27** | 16.27 | **59.51** |
| LR on bag of words | 29.55 | 70.45 | 32.23 | 55.66 |
| LR on bag of c3-grams | 30.68 | 61.36 | **32.39** | 56.84 |
| LR on doc2vec (our) | **42.20** | 64.77 | 31.29 | 55.90 |

Figure 2.3: Comprising of accuracy for doc2vec and baseline approaches

SVD, the singular-value-decomposition, as its proved in the previous researches. The SVD could greatly reduce the dimension they needed for the research, and it has shown the effect of many machine learning projects. The (Koren et al.2009) use SVD in the recommender systems, the author used the SVD to reduce the dimension of the vector space of movie features. The movie list is split to a train set and a test set, but the dimension of the movie matrix is too high and the sparse rate is 95.4% (4.6% non-zero data) is not acceptable for the recommendation engine, thus, in order to save time for the training, the researcher used the SVD to save most important information extracted from movie data After the implementation, the dimension of movie data is reduced to 700 and the F1 score dropped from 0.16 to 0.14, which is acceptable for the implementation.

Now we have discussed techniques of the n-gram model and SVD. Now we will discuss the machine learning techniques used in our system for candidate recommendation for jobs.

## 2.2.2   Machine learning approaches in recommender system

### Content-based recommendation

The research of (Paparrizos et al.2011) use the machine learning algorithms for job recommendation based on an employees job transaction. They first designed one baseline approach, which is built by predicting the most frequent job transaction in the user profile. Then they tested data using hybrid algorithm of nave bayes classifier. Finally,

| Dataset | Accuracy (%) | | |
|---|---|---|---|
| | Baseline | NB | Difference |
| I | 15.21 | 66.78 | 51.57 |
| II | 15.40 | 78.26 | 62.86 |
| III | 15.97 | 86.09 | 70.12 |

Figure 2.4: Comparison of accuracy with baseline approaches

they found that the machine learning approach improved more than 300% accuracy than baseline approaches.2.4

**Collaborative filtering recommendation**

Collaborative filtering is one of the most widely-used approaches in the recommender system. The researchers in (Lee et al.2001) have listed approaches i.e. users based and item based collaborative filtering approaches, and evaluation metrics i.e. root mean square error, precision and recall

Approaches like ((Li et al.2005), use a collaborative filtering approach for talent recommendation in e-commerce, they use candidate profile, job description, and candidate relationships for talent recommendation of job candidates. They successfully implemented a collaborative filtering talent recommendation with MAE lower than 0.2.

## 2.3   What I learned

By reviewing the literature which describes existing research in the area of recommender systems and machine learning, I have learned the following:

In order to build a successful recommender system, the first vital step is feature representation of the data. In the feature representation part, there are a range of techniques (i.e. N-grams, word embedding) that are capable of the feature representation. T h e  findings of research into using different learning algorithms are also very interesting.  We find that combining different machine learning algorithms and feature representation techniques could greatly increase the quality of recommendation achieved.

Another interesting result, based on the investigation of existing research, is that many researchers build recommender systems that follow a process of: (Data preprocessing ¿ Feature representation - ¿ Training the model using learning algorithms- ¿ recommend items) based on the research of (Paparrizos et al.2011) and (Ashok et

al.2009).

Thus, based on the research we have examined, we will select techniques of feature representation(e.g., bag-of-words, n-grams, word embedding etc. ) and machine learning algorithms to enrich our recommender system.

# Chapter 3

# Design and Methodology

## 3.1 Introduction

In this chapter, we will present the design and technical details of our recommender system. Initially, different learning types will be presented and discussed and we will select the most suitable learning type for our research aim. Then, based on the selected learning type, approaches to our recommender system will be selected. After selecting suitable approaches to the recommender system, data preprocessing, learning algorithms and feature representation techniques will be selected. Finally, the design pattern of the system will be presented to maintain the robustness and scalability of our recommender system.

## 3.2 Selecting Types of Learning for Recommender System

Machine Learning is a method of data analytics that aims to allow a computer to learn and make predictions regarding data based on an algorithm. Generally speaking, there are three types of machine learning:

Unsupervised learning: The machine learning algorithm tries to create a model purely based on the input. It looks for the inner patterns and structures in the data.

Supervised learning: The algorithm is presented with both input and the desired output. The algorithm aims to create a model that will best match the given inputs to the outputs. The goal of supervised machine learning is to build a concise model of the distribution of class labels in terms of predictor features (Kotsiantis, Zaharakis, and Pintelas, 2007).

Reinforcement learning: The algorithm is presented with a dynamic environment and a desired goal. It must learn which are the optimal actions it must undertake in order to achieve the target goal based on whatever state it is in.

In our research, supervised learning algorithms will be selected. The reason we select supervised learning is that unsupervised learning is based on unlabelled input and attempts to learn from the raw data, however, we prefer supervised learning for classification of candidates has we have training data with labelled job positions. For reinforcement learning, the environment of recommendation in this research is not dynamic, we use static data for recommendation and we do not have any dynamic status for our data thus there is no need for implementing reinforcement learning in the recommender system. The recommender system recommends candidates for jobs and we can use LinkedIn profile data as the input and job position as the output in supervised learning. In the following section we will introduce the approaches to recommender system based on our selection of learning type.

## 3.3    Selecting Recommender System Approaches

We have selected the learning type for our recommender system. Then we will discuss approaches in the recommender systems. The recommender system has mainly three popular approaches: content-based, collaborative filtering and hybrid approach.

The content-based and collaborative filtering approach for building approach will be selected for recommendation. The reason is that the content-based approach could tell us what feature are used in the recommendation. Thus we could evaluate the importance of features based on content-based approach. Another advantage for content-based approach is that it could solve the cold-start problem (i.e. the recommender system could not recommend similar items when adding a new item without any related information) in our recommender system because new items can be suggested without user ratings.

Collaborative filtering approach will also be selected because of following advantages like: Collaborative filtering approach work best when the user space is large (since that is their source of data). Content-based approach, however, needs technique for dimension reduction because content-based approach is more or less insensitive to user size. Meanwhile, adding collaborative filtering could fit our goal of the diversity of approaches for candidate/job recommendation.

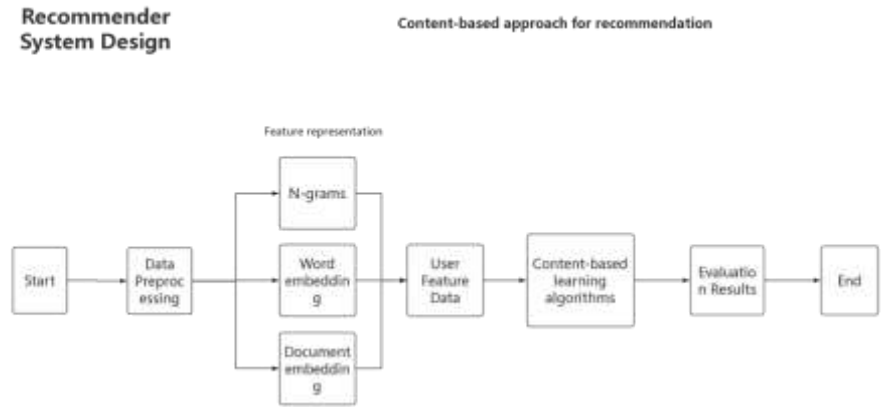

Figure 3.1: Comparison of accuracy with baseline approaches

The hybrid recommendation approach will not be considered in this research because our literature review hasnt covered the related researches. We will implement it in the future works.

Thus, based on the techniques used for collaborative filtering, we will select content-based and collaborative filtering approaches in our recommender system. We have developed recommender approaches from two aspects: content-based recommendation approaches using three different user feature representations, and collaborative filtering approaches. We will test the performance of different learning algorithms and find the best approach for the recommendation of the user profiles.

## 3.4  Design and Techniques of Content-based Recommender System

The high-level workflow of the recommender system is shown in Figure 3.1 We will introduce important components of our design:

Data preprocessing: In order to preprocess our data, we need to focus on two aspects: Data statistics and the design of a training/testing set. For data statistics, we will describe the details of the dataset and discuss what attributes of the user profile information could be used for different approaches in our recommender system. For generating training and testing set, we will discuss the feature engineering of the user data and then we will describe how training and testing data is generated for supervised learning.

Feature representation: We will implement 3 techniques to represent user feature.

First is the N-grams: the language model of using N-grams to analyze the contextual information in user profiles. We will demonstrate how N-grams works in our dataset and how to construct the feature data with our dataset. Then we will bring out word embedding and document embedding: the technique of reflecting word or document as feature vectors and how this technique will be implemented in our data. Finally, we will apply dimension reduction technique: the approach for dimension reduction in matrices formed by n-gram approaches. We will show why we need dimension reduction of feature vectors generated by N-grams and how to implement them in our approach. Learning algorithms for the recommendation are: Learning algorithms for content-based recommendation: we will discuss different supervised learning algorithms in machine learning for user profile recommendations, describe the reason for applying algorithms and the parameter settings of different algorithms.

### 3.4.1 Data analyzing and Preprocessing for Content-based Approach

In this section, we will present attributes of the data in the beginning. Then we will do data statistics and analysis.

Data cleaning for user profile data: The mix of number and words could be disturbing the performance of the learning algorithm, thus for attributes like work duration of work experience, the work experience is described as string of x year y month(s) in the original dataset, thus we use the regular expression to match the work experience string and convert them to numbers. Finally, we will normalize them to decimals between 0 and 1 in our data.

The dataset used in the research is collected from the popular professional social networking site LinkedIn, which consists of 158096 users. In our LinkedIn dataset, we have a CSV file with 158012 user profiles in the dataset containing background information of user talent. For every user, they have 67 attributes that can be categorized as following:

1. User id
2. Username
3. Connections number of user
4. Seven parts of work experience (current and past work experience, combined with job title, company name, company type, work duration, company location)
5. Four parts of educational backgrounds (university name, degree of education, major, end date of education, education details)
6. Skills
7. Languages

Those attributes could describe the general information of the user profile with their talents.

Analyzing of attributes: we found some attributes with insignificant information, pure numbers or combination of number and words: user id (section1), username (section2), connection number of the user (section 3), work duration of work experience and graduation year of the education background (section 5). In this research we will remove username, con- nection number of user and graduation year of the school in the data we use because for username and connection number, we could neither dig more semantic nor contextual data those attributes. Remaining attributes will be analyzed through data statistic.

Data cleaning for user profile data: Before the statistic of data, we need to implement the data cleaning to avoid the obstruct of messy data. The mix of number and words could be disturbing the performance of the learning algorithm, thus for attributes like work duration of work experience, the work experience is described as string of 'x year y month(s)' in the original dataset, thus we use the regular expression to match the work experience string and convert them to numbers. Finally we will normalize them to decimals between 0 and 1 in our data.

Data analysis for job titles: In our dataset, the top 15 job titles in the dataset are shown in Figure 3.2

We need to analyze job title based on following reasons: For online recruitment websites, the foremost attribute in a job is the job title and most of users in job websites find jobs based on job titles. Thus we could set user data for the input of recommender system and set the prediction of job title as output of our system. Based on the statistic of job titles, we selected six different job titles for recommendation LinkedIn user to job (account manager, software engineer, research assistant, project manager, process engineer, consultant). The selected job titles are from different areas and some job titles have clear statement (software engineer), some are not (e.g., consultant). Thus, we could test the performance of different approaches thoroughly by using different job titles.

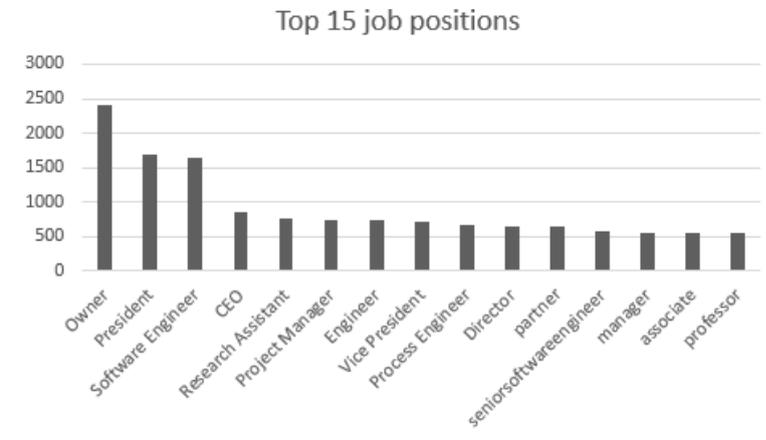

Figure 3.2: Top 15 job titles in the dataset

Work experience statistic: we want to learn the work experience in our LinkedIn users. Thus we count the number of work experience per user. According to the statistical data of our user profiles. The number of effect value in seven work experience is shown in Figure 3.3

In this diagram, recent means the current job position of user, and past1-6 means the past work experiences of user. Thus for every user, the average number of effective work experience number per user is 2.19, thus for every user profile there is at least 2 work experience, we could use this feature in our recommendation system.

Education background of the user: We want to investigate education in LinkedIn user profile. The Figure 3.4 showed the total number of education experience in four education experience. Thus based on the statistic data, the average education background per user is 1.76, which can be used for talent recommendation.

This work also did the statistics on the averaged number of skills per user listed on his LinkedIn profile. Skills in the user profile are given by a long sentence separated by commas, thus we split skills by comma and obtained the total number of skills, and calculated the averaged number of skills per user: 19.72. The number shows the skill column contains plenty of user information.

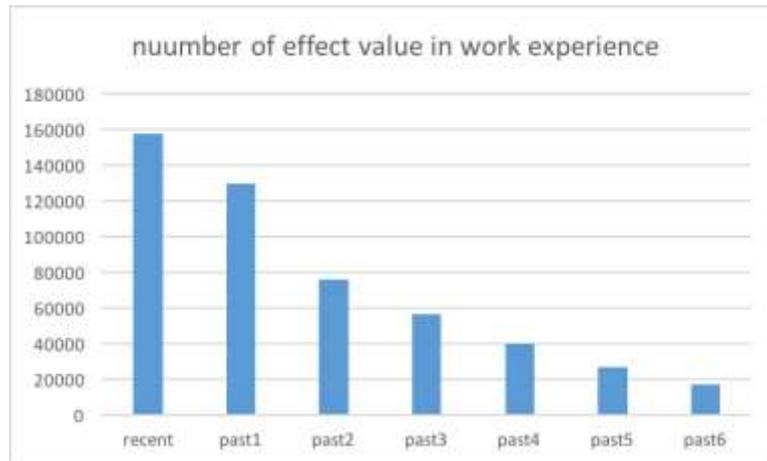

Figure 3.3: Number of work experience

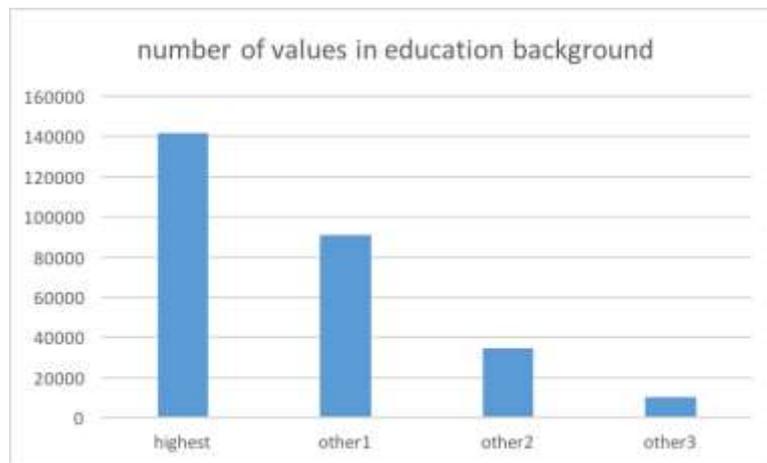

Figure 3.4: Number of education background

Thus, on average, every user has at least 2 work experience, 1 education background, and 19 skills, we could use those attributes for talent recommendation.

### 3.4.2 Building Training set and testing set and Setting label data for Content-based recommendation

Based on the analyzing and prepossessing of the data in the section3.5, we decided to use the section 4,5,6,7 (i.e. seven work experience, four education background, skills and languages attributes) of the user data in the section three, part of LinkedIn user profile data.

Supervised learning for job recommendation is chosen in this research and our research aim is to identify suitable approaches for candidate recommendation for jobs. Thus, for a specific job title, we could judge whether a user is suitable for given job title by setting the LinkedIn user profile data as input, and the output is the binary value of the prediction of user job title (i.e. If the job title is the chosen job title, the value will be 1, otherwise the value will be 0). Thus we selected the job title of the first work experience in the seven work experience as labels.

In our project, we will select 1000 users based on chosen job title. We will randomly select 500 users whose label is 1 (i.e. job title of first work experience is same with chosen job title) and 500 users whose label is 0(i.e. first work experience job title is not the chosen job title). The process of generating train set and test set is shown in Figure3.5

### 3.4.3 Techniques for Feature Representation of Content-based Recommender System

In this section we will contain 3 techniques of feature representation: N-grams model, word embedding and document embedding and 8 learning algorithms in the content-based recommendation approach. We will present the details for those techniques and how we implement them in our recommender system.

**Feature representation using N-grams**

N-grams is widely used in probability, communication theory, computational linguistics (for instance, statistical natural language processing), computational biology (for instance, biological sequence analysis), and data compression. The advantage of using N-gram models (and algorithms that use them) are simplicity and scalability with

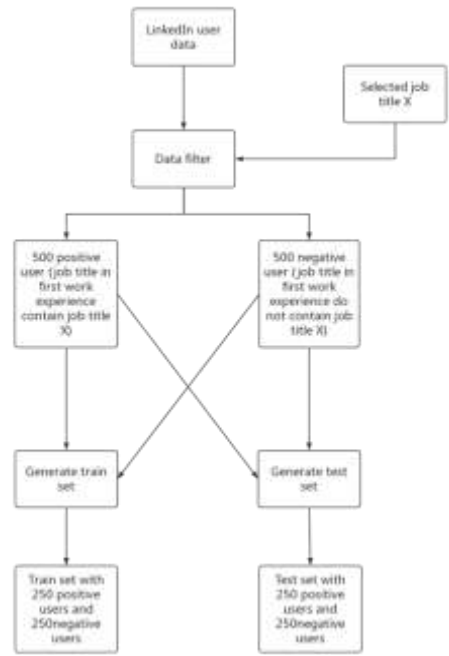

Figure 3.5: Workflow of train set and test set generation

larger n, a model can store more context with a well-understood space-time tradeoff, enabling small experiments to scale up efficiently. The word representation of the n-grams is one-hot encoding in the user profile, which means words in the data will be shown as the values in the sentence-word vectors if the sentence contains this word (i.e. if the sentence contains the word the value will be 1, otherwise 0). We can see N-grams could use context information in the user profile based on setting the N value (i.e. Length of tokens) in N-grams. To extract context information, we propose to use the n-gram to extract the context information of users. Concretely, for a candidate, his profile will be tokenized (e.g. for trigram(length of token is 3), the text software engineer work in microsoft will be tokenized as list of software engineer work, engineer work in, work in microsoft) By doing so, the context information can be saved. In the experiments of this research project, following approaches are considered:

We can see n-grams could use context information in the user profile based on using different n-grams. To identify experts we propose to use the n-gram to extract the context information of users. Concretely, for an expert candidate, his user profile will be split into a collection of n words combinations (e.g. for 2-gram split text senior

**Word-level unigrams**

| Text | Token Sequence | Token Value |
|------|----------------|-------------|
| consult three years deloitte | 1 | consult |
| consult three years deloitte | 2 | three |
| consult three years deloitte | 3 | years |
| consult three years deloitte | 4 | deloitte |

Figure 3.6: Unigram model

**Word-level bigrams**

| | Token Sequence | Token Value |
|------|----------------|-------------|
| consult three years deloitte | 1 | consult three |
| consult three years deloitte | 2 | three years |
| consult three years deloitte | 3 | years deloitte |

Figure 3.7: Bigram model

software engineer as senior software,software engineer) thus every n-gram text will be related to its prior text. By doing so, the context information can be saved. We select a sentence 'consult three years deloitte' from the user profile and experiments from N-grams are considered:

   1.Unigram, i.e. bag-of-words the length of token is 1, the approach is illustrated in Figure3.6

   2.Bigram (2-gram): the length of token is 2, the diagram is shown in Figure3.7

   3. Trigram (3-gram): the length of token is 3, the diagram is shown in Figure3.8

## Feature Representation Using Word/document embedding model

We already finished feature representation using the n-gram for the content-based recommendation and contained partially contextual and semantic information in every user profile. Now we need to investigate more approaches for extracting semantic in-

**Word-level trigrams**

| Text | Token Sequence | Token Value |
|------|----------------|-------------|
| consult three years deloitte | 1 | consult three years |
| consult three years deloitte | 2 | three years deloitte |

Figure 3.8: Trigram model

formation in user profiles. Different from N-grams, the word embedding and document embedding using tools of word2vec and doc2vec to forming word vector/ document vector to represent the semantic information of words in the the user profile. The word/document vectors usually contain the semantic information of the paragraph or the entire dataset.

Word2vec is a well-known tool that builds word embedding using the hidden neural network for prediction of words. Based on the research by Omer Levy (Goldberg & Levy2014), the advantage of using word2vec is that it works well on many semantic tasks like word similarity and word analogy, another advantage is that word2vec could use many pre-trained models such as the Glove and Google negative 300 and they could provide fairly good performance.

In order to extract the semantic information using word2vec, we need to tokenize each user profile into a list of tokens, then we could use pre-trained word embedding models (such as Glove, Google negative300) to convert words in the user profiles to word vectors, after that, we could calculate the vector of the user profile by calculating the average of the sum of word vectors. Finally, we could use user profile vectors in our learning algorithms. The vector of user profile contains the semantic information of words in the data. The process of forming user profile vector is shown in Figure3.9.

The doc2vec is the further development of the word2vec, in the input layer of doc2vec, the doc2vec adding a new paragraph vector on the basis of word2vec and it will produce word vector for each word in the document and document vector for each document by adding a list of documents as training data.The Figure3.10shown how to extract semantic information inside documents.

To implement document embedding with LinkedIn user profile data, we need to train the user profile dataset with doc2vec and get word and document vectors. The document vectors contain the context information of paragraphs in the user profiles. Then we set every document as a input, the doc2vec will transfer the input document to a vector based on existing word and document vectors in the model.

## Approaches for Dimension Reduction in Content-based Recommendation

The N-gram model could extract context information in the LinkedIn user profile data. However, feature vectors formed by N-grams are usually with high dimension shown in

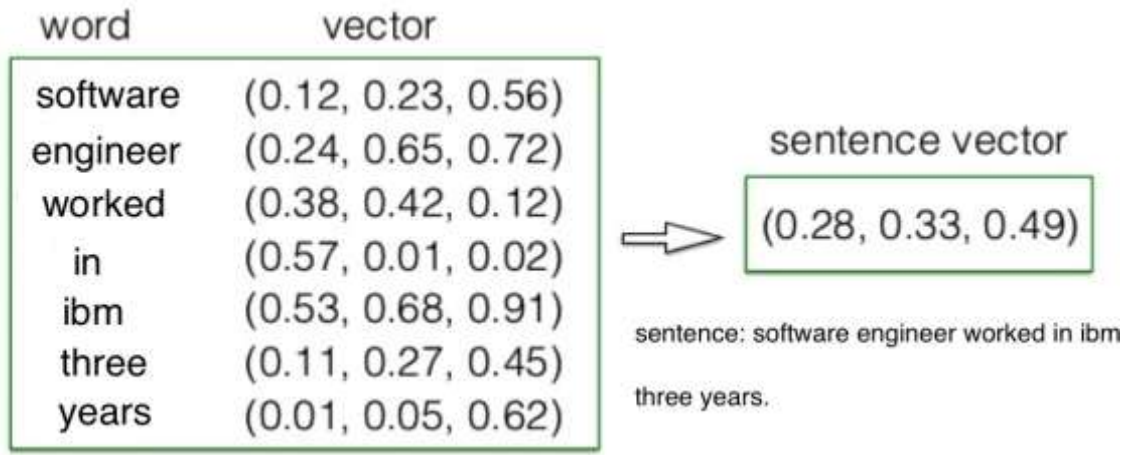

Figure 3.9: Forming user profile vector by word2vec

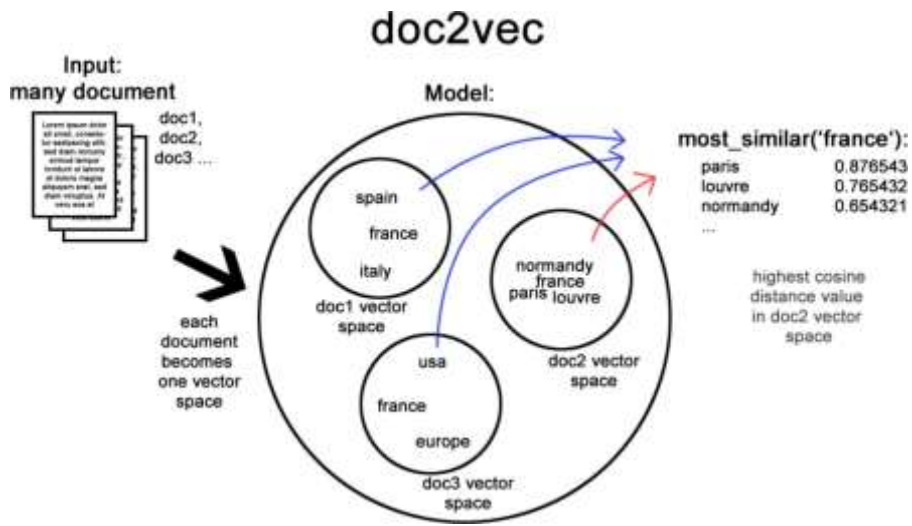

Figure 3.10: Doc2vec model

the table 3.1

Many problems caused by sparse matrices: The sparse matrix will increase the training time of the user profile data, which can be seen in table 3.2 and a long training time is unacceptable for many recommender systems. A long training time for the recommender system could result in terrible user experience and cause terrible user experience leading to the economically lost. Thus, lowering the dimension of user feature representation (form a dense data) is crucial for our recommender system. For reducing the training time, we need to implement the singular value decomposition, which is an approach for dimension reduction of feature data. In the previous section, we found that feature representation using N-grams will form a high dimension of feature vectors, which is very sparse.

| | Software engineer | Logistic regression CV | SVM Linear SVC | SVM NuSVC | SVM SVC | Naive Bayes |
|---|---|---|---|---|---|---|
| Dimension value | 10963 | 9847 | 22456 | 16728 | 26365 | 19851 |

Table 3.1: Dimension value of generated feature vectors by Bigram

| | Logistic regression | Logistic regression CV | SVM Linear SVC | SVM NuSVC | SVM SVC | Naive Bayes | Decision Tree | Random forest |
|---|---|---|---|---|---|---|---|---|
| Time costing | 0.5495 | 10.1976 | 0.5268 | 11.2280 | 11.4863 | 0.8090 | 1.3946 | 0.4606 |

Table 3.2: Average precision for every learning algorithms

The SVD algorithms can be used in the dimension reduction of vector space, which can reshape the vectors and contains the most important features inside the user profile and give faster results.

The explanation of the singular value decomposition is shown as follows: In theory, the singular value decomposition is to decompose the matrix into the sum of several rank 1 matrix, then save the value with the high singular value in the formula inside the matrix and remove value with the low singular value inside the formula. Thus, the generated new matrix is smaller than the original matrix and it could reduce the training time of the model. In our example, we will use a picture with pixel 450*333. We form a matrix A with 450 row x 333 column based on the pixel of the picture.

Then we do the singular value decomposition to the matrix and get the result: The formula of the SVD is shown in Formula 3.1:

$$A = \sigma_1 u_1 v_1 + \sigma_2 u_2 v_2 + \dots\dots + \sigma_n u_n v_n \tag{3.1}$$

where $\sigma$ is singular value, $u$ and $v$ are the column vectors.

If we only contain the first value of the formula, which contains a highest singular value, the formula is 3.2

$$A_1 = \sigma_1 u_1 v_1 \tag{3.2}$$

We can see the picture is still blurry in figure 3.11 because we captured few features of the picture.

If we contain the first 5 values of the formula, which is shown as following:

$$A = \sigma_1 u_1 v_1 + \sigma_2 u_2 v_2 + \sigma_3 u_3 v_3 + \sigma_4 u_4 v_4 + \sigma_5 u_5 v_5 \tag{3.3}$$

We could see the outline of the figure in 3.12

If we contain the top 50 values of the formula, the picture is: We can see the new picture is very closed to original picture in figure 3.13

The new picture that contains the top 50 features has the pixel of 784*50 = 39200, which is much smaller than the original picture with the pixel of 450*333 = 149850.

The reduced matrix using SVD could greatly reduce the training time of the model, the result will be shown in the next chapter.

Thus, in our project, the feature vectors of LinkedIn users will be processed through singular value decomposition from high dimension to 50 dimension, preserve essential features while removing insignificant features.

## Learning algorithms for Content-based Recommendation

Different algorithms will be tested on those vector spaces. We will present eight different learning algorithms and demonstrate how we implement them:

1. Logistic Regression
2. Logistic Regression CV
3. SVM Linear SVC
4. SVM NuSVC

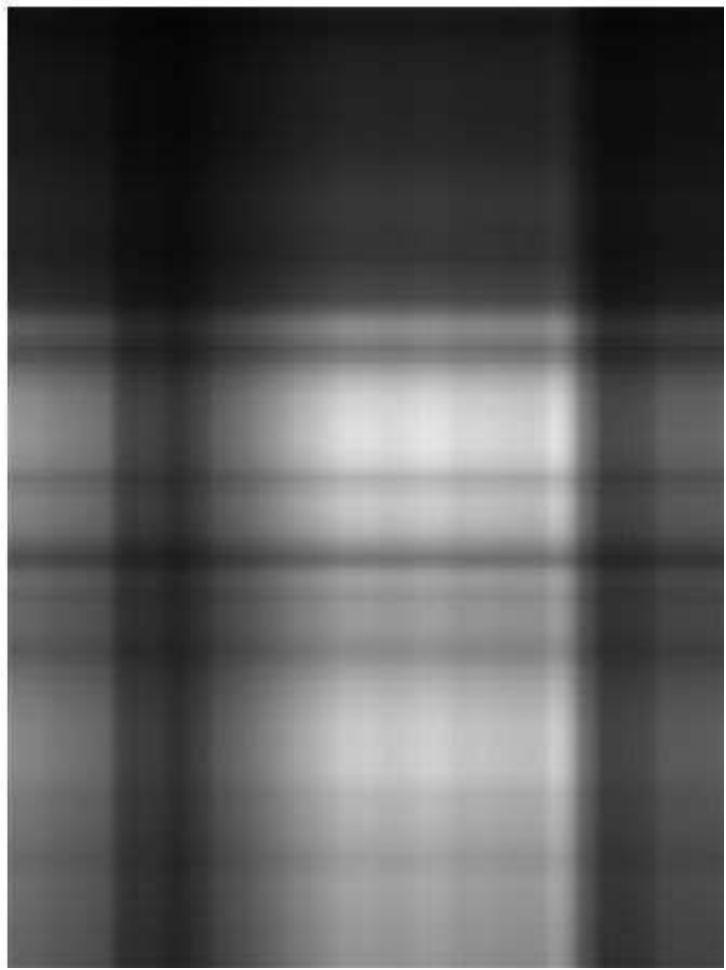

Figure 3.11: Figure that only consider top 1 value in SVD formula

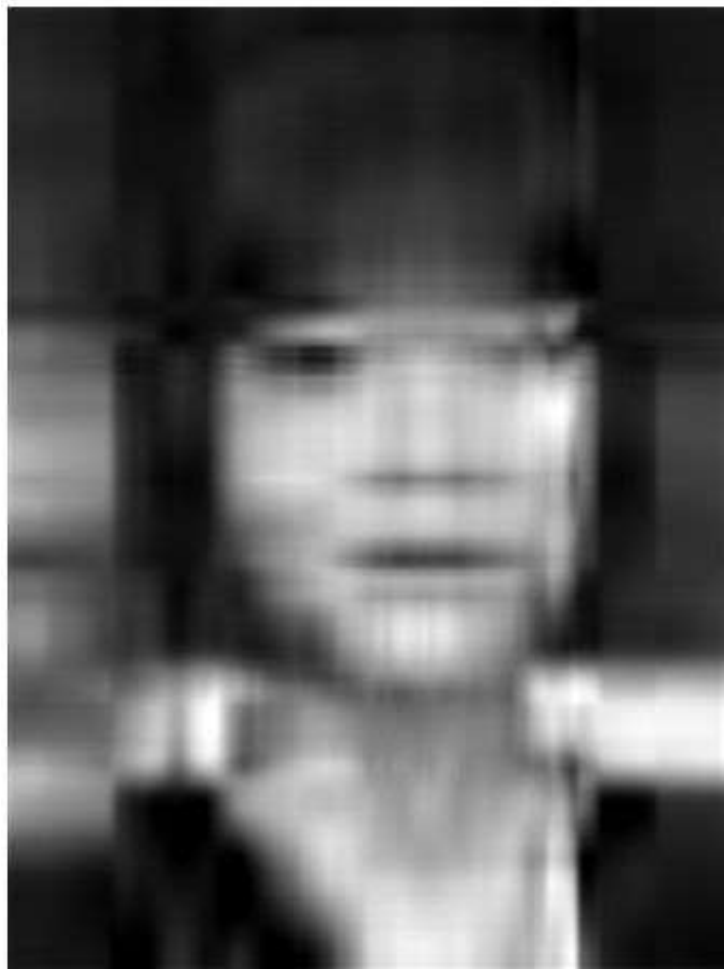

Figure 3.12: Figure that only consider top 5 value in SVD formula

5.SVM SVC

6.Naive Bayes

7.Decision Tree

8.Random forest

**1) Logistic Regression**

Logistic regression is a supervised learning algorithm which could be used in both

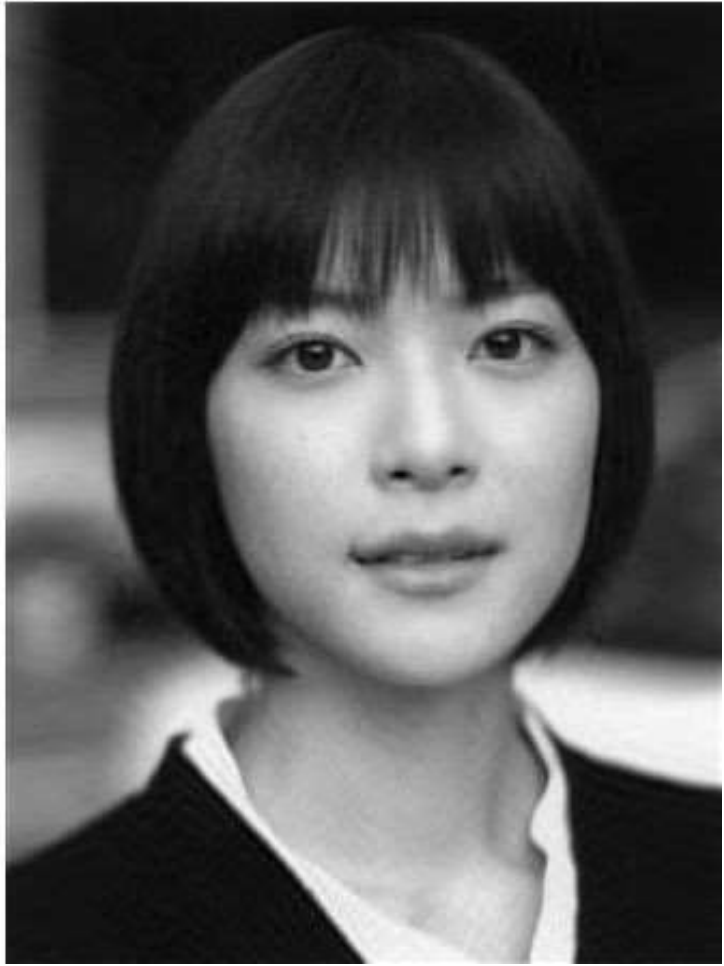

Figure 3.13: Figure that consider top 50 value in SVD formula

classification and regression. In this research, given a certain factors, logistic regression is used to predict an outcome which has two values such as 0 or 1, pass or fail, yes or no by calculated probabilities. Probabilities are estimated using logistic/sigmoid function. The graph of sigmoid function is an S curve. Figure of the Sigmod function is 3.14

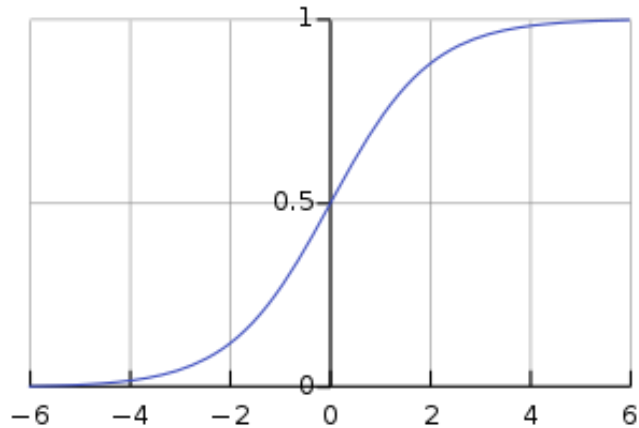

Figure 3.14: Figure of the sigmod function

The formula is shown as following:

$$f(t) = \frac{1}{1 + e^{-t}} \tag{3.4}$$

where

$$t = w_0 + w_1 x_1 + w_2 x_2 + \ldots + w_n x_n \tag{3.5}$$

.

Here $w_1$, $w_2 \ldots w_n$ are the regression co-efficients of the model and are calculated by Maximum Likelihood Estimation and $x_1$, $x_2 \ldots x_n$ are variables of formed feature vectors of LinkedIn user data. f(t) calculates the probability of the binary outcome and using the probabilities we classify the given LinkedIn user data into one of the two categories. The logistic regression CV is Logistic regression using k-fold cross validation.

**2) SVM**

Support vector machine (SVM) is a supervised learning algorithm that could be used for regression and classifying. In this project we will use SVM for classification. The reason why we implement SVM algorithm has several advantages like: SVMs can produce accurate and robust classification results on a sound theoretical basis, even when input data are non-monotone and non-linearly separable. SVM have the benefit of good performing in the high dimension spaces such as the feature data in our project. For these reasons, SVMs are regarded as a useful tool for effectively classifying techniques.

The SVM algorithm works by plotting data as points in the n-dimensional space (Where n is the feature number in the LinkedIn data profile). Then SVM will find a hyperplane which could best split two datasets. The figure is shown in figure 3.15.

In our project, we use SVM SVC (C-Support Vector Classification), SVM NuSVC (Nu-Support Vector Classification), SVM Linear SVC (Linear support Vector Classification) to test our data.

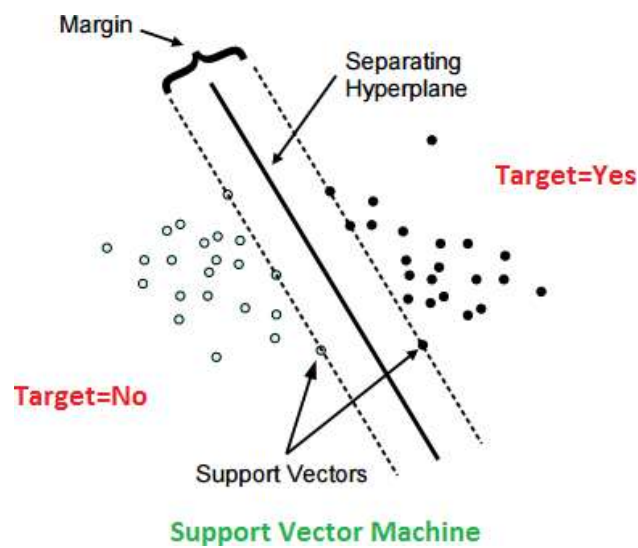

Figure 3.15: Theory of SVM

### 3) Naive Bayes

We decide to implement the Naive Bayesian classifier based on following reasons: it is easy to build, with no complicated iterative parameter estimation which makes it particularly useful for very large datasets. Despite its simplicity, the Naive Bayes classifier often does surprisingly well and is widely used because it often outperforms more sophisticated classification methods.

In our research, we will set the user profile as the input and then the classifier will calculate the conditional probability of features for classification and then classify LinkedIn user data to suitable job titles.

### 4) Decision Tree

The decision tree is supervised learning algorithm which builds classification or regression models in the form of a tree structure. It breaks down a dataset into smaller and smaller subsets while at the same time an associated decision tree is incrementally developed. The final result is a tree with decision nodes and leaf nodes. A decision node (e.g., Outlook) has two or more branches (e.g., Sunny, Overcast and Rainy). Leaf node (e.g., Play) represents a classification or decision. The topmost decision node in a tree which corresponds to the best predictor called root node. The structure of model is shown in Figure 3.16. The advantage of decision tree is: Easy to understand and can handle both categorical and numerical data.

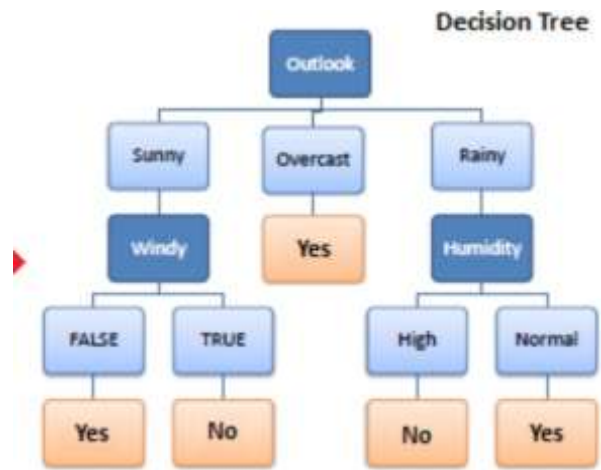

Figure 3.16: Theory of decision tree

### 5) Random Forest

Random forest is an ensemble machine learning technique used to perform classification. Ensembles combine multiple hypotheses to form a (hopefully) better hypothesis. The main principle behind ensemble methods is that a group of weak learners can come together to form a strong learner. Random forest uses a multitude of decision trees (weak learners) to perform its classification, which is shown in Figure 3.17. The random forest could average or combine the results of different decision trees helps to overcome the problem of overfitting. We will put the user profile data as input and output the target value (i.e. prediction of the job title for LinkedIn users). In this re-search , the feature data will be the input of the learning algorithm and the prediction of job title will be the output.

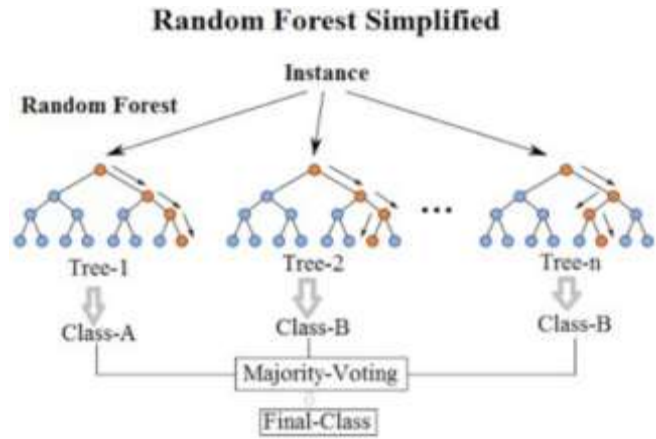

Figure 3.17: Theory of decision tree

## 3.5   Design and Methodology of the Collaborative Filtering Approach

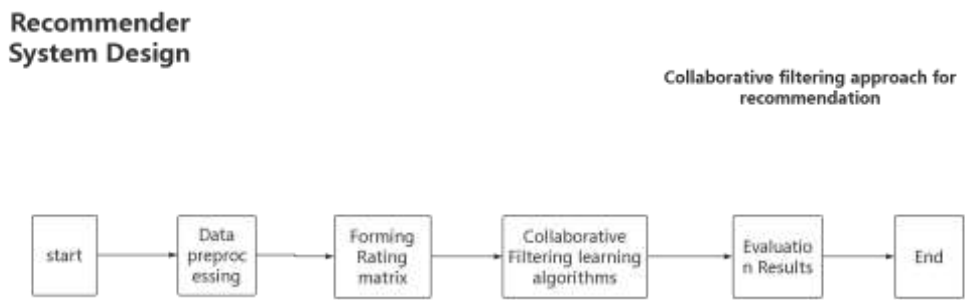

Figure 3.18: Design of Collaborative filtering approach

The design of collaborative filtering is shown in figure 3.18. We will introduce main components in the collaborative filtering approach:

Data preprocessing: In the data preprocessing part we will discuss what attributes will be suitable for collaborative filtering approaches.

Forming matrix for learning algorithm: Process of forming user rating matrix will be presented and we will illustrate how we form the rating value for collaborative learning algorithm.

### 3.5.1 Data preprocessing for Collaborative Filtering Recommendation

For the collaborative filtering approach, we usually need to form a matrix based on our goal for the recommendation. For example, for a recommendation in e-commerce, we usually need to form user-item matrix to find similarity between customers. For a similar book recommendation, we usually need to build item-item matrix to show the inner connection between books.

Thus in our collaborative filtering approach, we decided to form the matrix of user-workexperience-rating matrix (i.e. user as row of matrix, work experience attribute except the first work experience as columns of matrix and value in cells is the rating of work experience, which will be discussed thoroughly in the feature representation section) for collaborative filtering learning algorithms. The reason why we implement the strategy is that the collaborative filtering approach could fit our research aim (i.e. suitable approach for candidate recommendation for jobs). Thus we consider using work experience in the collaborative filtering approach. In order to build data for supervised learning, the job title of first work experience will be the output of learning and user-workexperience-rating matrix will be the output.

### 3.5.2 Building Matrix for Collaborative Filtering Approach

We will build a user-workexperience-rating matrix based on the user profile data. The work experience (Columns of matrix) is six work experiences of a user mentioned in section 3.3 and we calculate the rating based on the following rules: For a specific given job title (i.e. software engineer), we will calculate the rating based on following rules:

1. If the job title in the work experience is same as the given job title and the work year is more or equal than 3 years, the rating value is 2.

2. If the job title in the work experience is the given job title and the work year is less than 3 years, the rating value is 1.

3. If the job title in the work experience is not the given job title, the rating value is -1.

Thus we could form a matrix of user-workexperience-rating in figure3.13.19

| | work_exp1 | work_exp2 | work_exp3 | work |
|---|---|---|---|---|
| 143872 | 0 | 0 | 1 | |
| 562381 | 0 | 0 | 0 | |
| 39423 | 1 | 0 | 1 | |
| 760783 | 0 | 0 | 0 | |

Figure 3.19: Snippet of the matrix

### 3.5.3 Learning Algorithms for Collaborative Filtering

We will test 2 different collaborative filtering algorithms in the research: SVDPlusPlus Recommender ASVDPlusPlus.

**SVDplusplus Leanring Algorithm**

In an easier way to explain terminology, SVD (i.e. singular value decomposition) finds the latent factors associated with some matrix. For example, in recommender systems, the user-rating matrix of movies after an SVD, will decompose into matrices that represents latent user-user features and item-item features e.g. same gender user, same age-group of users, action movies etc. and many other latent factors involved in the rating behavior that is not apparent from the user-rating matrix.3.19Based

$$X \begin{pmatrix} x_{11} & x_{12} & x_{1n} \\ x_{21} & x_{22} & \\ \vdots & \vdots & \ddots & \\ x_{m1} & & x_{mn} \end{pmatrix}_{m \times n} = U \begin{pmatrix} u_{11} & \cdots & u_{1r} \\ \vdots & \ddots & \\ u_{m1} & & u_{mr} \end{pmatrix}_{m \times r} S \begin{pmatrix} s_{11} & 0 & \cdots \\ 0 & \ddots & \\ \vdots & & s_{rr} \end{pmatrix}_{r \times r} V^T \begin{pmatrix} v_{11} & \cdots & v_{1n} \\ \vdots & \ddots & \\ v_{r1} & & v_{rn} \end{pmatrix}_{r \times n}$$

Figure 3.20: Theory of SVDplusplus

on the research of (Koren2008), the SVDplusplus (i.e. singular value decomposition plus plus) incorporate the implicit feedback of the rating data (i.e. Possible preference for a specific job title such as data scientist for a user with work experience such as software engineer) by generating feature matrix by matrix factorization. In this research, the rating value in matrix will be selected as the input and the output of matrix factorization will be feature matrix with lower dimension. Thus, in order to find out the implicit information of user feature

**ASVDplusplus Learning Algorithm**

The ASVDplusplus (Asymmetric SVDplusplus) is the simplified version of the SVD-plusplus learning algorithm because of reduced parameter in the feature extraction function. In this research, the rating value in matrix will be selected as the input and the output of matrix factorization will be feature matrix with lower dimension. Thus, in order to find out the implicit information of user feature, we will implement ASVDplusplus algorithm in the collaborative approach.

# 3.6 Design pattern

## 3.6.1 Introduction

Organized code style and structure is essential for software engineering. Well-organized code will perform better in usability, scalability, and robustness of the software system.

We will develop the system based on the following concepts in order to make the recommender engine more robust and scalable.

## 3.6.2 Object oriented programming

In our system, we developed the object-oriented advantages of the object-oriented programming:

Object-Oriented Programming has great advantages over other programming styles: Code Reuse and Recycling: Objects created for Object-Oriented Programs can easily be reused in other programs.

Encapsulation (part 1): Once an Object is created, knowledge of its implementation is not necessary for its use. In older programs, coders needed to understand the details of a piece of code before using it (in this or another program).

Encapsulation (part 2): Objects have the ability to hide certain parts of themselves from programmers. This prevents programmers from tampering with values they shouldnt. Additionally, the object controls how one interacts with it, preventing other kinds of errors. For example, a programmer (or another program) cannot set the width of a window to -400. Design Benefits: Large programs are very difficult to write. Object-Oriented Programs force designers to go through an extensive planning phase, which makes for better designs with fewer flaws. In addition, once a program reaches a certain size, Object Oriented Programs are actually easier to program than non-Object Oriented ones.

Software Maintenance: Programs are not disposable. Legacy code must be dealt with on a daily basis, either to be improved upon (for a new version of an existing piece of software) or made to work with newer computers and software. An Object-Oriented Program is much easier to modify and maintain than a non-Object Oriented Program. So although a lot of work is spent before the program is written, less work is needed to maintain it over time. In our system, we will form a class named LinkedIn data, the class will have 67 attributes corresponded to 67 columns in the dataset. The procedure of generating data is 1. Initiating the data: every row in our dataset is a user profile and every column is the attribute of the user, thus we generate the LinkedIn data class while reading every line of the CSV file. Then we generate the data into the list of the class.

Based on the analysis of our data, the organization of our data is shown in Figure 3.21:

As a result, the attributes of the user profile class in Python is shown as follows:

```
class LinkedInData (object):

    def _init__(self, id, name, connections, title, org_summary,
    org_detail, duration, location, description,
                past job title1, past job org summary1,
    past_job_org_detail1, past job duration1, past job location1,
                past job description1,
                past job title2, past job org summary2,
                ........................
past_job_org_detail6, past job duration6, past job location6,
                past job description6,
                highestLevel_universityName, highestLevel degree,
```

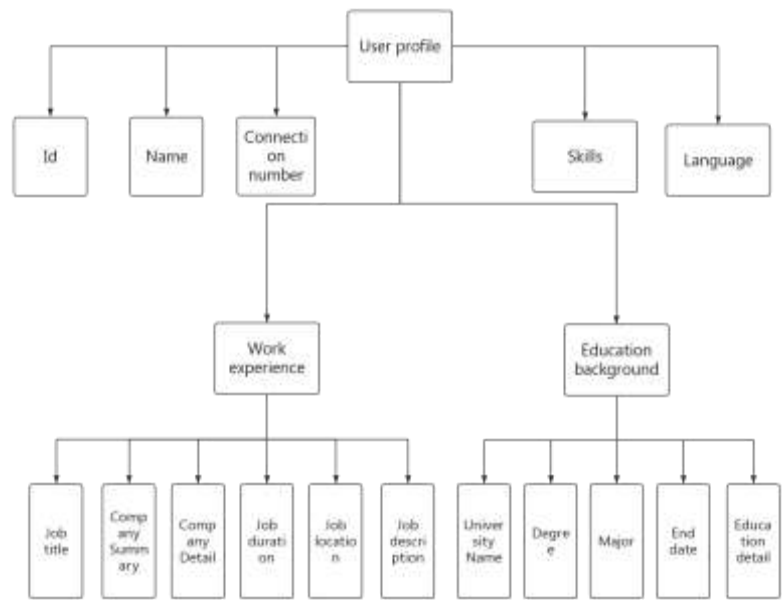

Figure 3.21: Organization of user profile class

```
           highestLevel_major,
11                    highestLevel_endDate , highestLevel_detail ,
12                    otherLevel_universityName1 ,   otherLevel_degree1 ,
     otherLevel_major1 , otherLevel_endDate1 ,
13                    otherLevel_detail1 ,
14                    otherLevel_universityName2 ,   otherLevel_degree2 ,
     otherLevel_major2 , otherLevel_endDate2 ,
15                    otherLevel_detail2 ,
16                    otherLevel_universityName3 ,   otherLevel_degree3 ,
     otherLevel_major3 , otherLevel_endDate3 ,
17                    otherLevel_detail3 ,
18                    skills , languages ):
19        self.id=id
20        self.name = name
21        self.connections = connections
22        self.title= title
23        self.org_summary = org_summary
24        self.org_detail = org_detail
25        self.duration = duration
26        self.location = location
27        self.description=description
```

```
28          self.pastjobtitle1,self.pastjoborgsummary1, self.
      pastjoborgdetail1,\
29          self.past_job_duration1, self.past_job_location1, \
30          self.past_job_description1 = past_job_title1,
      past_job_org_summary1 \
31          , past_job_org_detail1, past_job_duration1,
      past_job_location1, past_job_description1
32          self.past_job_title2, self.past_job_org_summary2, self.
      pastjoborgdetail2,\
33          self.past_job_duration2, self.past_job_location2, \
34          self.past_job_description2 = past_job_title2,
      pastjoborgsummary2 \
35          , past_job_org_detail2, past_job_duration2,
      past_job_location2, past_job_description2
36          ...................
37          self.past_job_title6, self.past_job_org_summary6, self.
      pastjoborgdetail6,\
38          self.past_job_duration6, self.past_job_location6, \
39          self.past_job_description6 = past_job_title6,
      pastjoborgsummary6 \
40          , past_job_org_detail6, past_job_duration6,
      past_job_location6, past_job_description6
41
42          self.highestLevel_universityName, self.highestLevel_degree, self.
      highestLevel_major, \
43          self.highestLevel_endDate, self.highestLevel_detail =
      highestLevel_universityName, highestLevel_degree \
44          , highestLevelmajor, highestLevelendDate,
      highestLeveldetail
45
46          self.otherLevel_universityName1, self.otherLevel_degree1, self.
      otherLevelmajor1,self.otherLevelendDate1,\
47          self.otherLevel_detail1 = otherLevel_universityName1,
      otherLevel_degree1, otherLevel_major1 \
48          , otherLevelendDate1,otherLeveldetail1
49          ....................
50          self.otherLevel_universityName3, self.otherLevel_degree3, self.
      otherLevel_major3, self.otherLevel_endDate3, \
51          self.otherLeveldetail3 = otherLeveluniversityName3,
```

```
      otherLeveldegree3, otherLevelmajor3 \
52            , otherLevel endDate3, otherLevel detail3
53        self.skills=skills
54        self.languages=languages
```

Listing 3.1: LinkedIndata.class

# Chapter 4

# Evaluation and Discussion of Research Results

To evaluate the performance of various recommendation approaches on the constructed LinkedIn dataset, We will first introduce the metric for evaluation of our approaches. Then we look at different aspects of the performance of these approaches: training time, performance with different sizes of training data, different learning algorithms and different feature representation of data.

## 4.1   Evaluation metrics for recommender system

**Precision and Recall**

For better evaluation of different approaches, we will calculate the precision, recall and accuracy value of different approaches. The precision formula is shown as follows:

$$Precision = \frac{T\ P}{TP + FP} \tag{4.1}$$

The recall formula is shown as follows:

$$Recall = \frac{T\,P}{T\,P + F\,N}$$

Where, TP is true positive, FP is false positive, TN is true negative, FN is false negative. The mean of those words are:

True positive(TP): The number of correct prediction of relevant users in the LinkedIn profile (e.g., predicting the relevant user as relevant).

True negative(TN): The number of the correct prediction of non-relevant users (e.g., predicting non-relevant users as non-relevant).

False positive(FP): The number of the false prediction of non-relevant users (e.g., predicting non-relevant users as relevant).

False negative(FN): The number of the false prediction of relevant users (e.g., predicting relevant users as non-relevant)

**Average precision and recall:**

Considering our research aim, we will calculate average precision and recall of every approach by average the sum of precision value and recall value for all select job titles.

# 4.2 Experimental results and analysis

In this section, we will present experimental results and result analysis. We will split the evaluation in two main section: Evaluation for content-based approaches and evaluation for collaborative filtering approaches. In the evaluation of content-based approach, results from be presented based on different feature representation, then in every feature representation results, the performance of different learning algorithms for different job title will be presented. In the evaluation of collaborative filtering approach, the results of different learning algorithms for different job title will be discussed.

## 4.2.1 Result Evaluation in Content-based Approach

### N-grams for Feature Representation in the Content-based Approach

We have results of 3 different N-grams (Unigram, Bigram, Trigram) for feature representation and 8 different learning algorithms (Logistic regression, Logistic regression CV, SVM Linear SVC, SVM NuSVC, SVM SVC, Nave Bayes, Decision Tree, Random forest, ) and 4 different parts of user data (All data, user work experience of the data, education background of user data and skills of user data). In total we have 96 results. In order to identify approaches that could support talent recommendation, we will evaluate them by calculating the average precision of all job titles for every combination of feature representation, learning algorithm and user data. Then we can find the best combination of feature representation, learning algorithm and user data which could fit most job titles.

1. Approaches using all data: In this part, I use all user data(i.e. including work experience, education background and skills in the chapter 3.2.2) We could observed from the table 4.1 that the approach with highest average precision 0.7975 is the combination of (Trigram + Random forest learning algorithm + all data)

|  | Logistic regres­sion | Logistic regres­sion CV | SVM Linear SVC | SVM NuSVC | SVM SVC | Naive Bayes | Decision Tree | Random forest |
|---|---|---|---|---|---|---|---|---|
| Unigram | 0.7429 | 0.7448 | 0.7354 | 0.6727 | 0.7157 | 0.6444 | 0.7333 | 0.7440 |
| Bigram | 0.7299 | 0.7108 | 0.7524 | 0.7188 | 0.7219 | 0.5842 | 0.7628 | 0.7655 |
| Trigram | 0.7081 | 0.6751 | 0.7817 | 0.7053 | 0.7095 | 0.5497 | 0.7776 | 0.7975 |

Table 4.1: Average precision for all job positions of using all data

|  | Logistic regres­sion | Logistic regres­sion CV | SVM Linear SVC | SVM NuSVC | SVM SVC | Naive Bayes | Decision Tree | Random forest |
|---|---|---|---|---|---|---|---|---|
| Unigram | 0.6641 | 0.6522 | 0.6374 | 0.7445 | 0.7137 | 0.7656 | 0.5742 | 0.5448 |
| Bigram | 0.5026 | 0.5392 | 0.392 | 0.5852 | 0.5453 | 0.774 | 0.4597 | 0.4153 |
| Trigram | 0.3516 | 0.4782 | 0.168 | 0.4192 | 0.4038 | 0.7574 | 0.414 | 0.3457 |

Table 4.2: Average recall for all job positions of using all data

2. Approaches using work experience data: In this part, I use all user data(i.e. including work experience, education background and skills in the chapter 3.2.2) We could observed from the table 4.3 that the approach with highest average precision 0.7606 is the combination of (Bigram + Random forest learning algorithm + all data)

3. Approaches using work education background: In this part, I use education background data(i.e. education background data used in the chapter 3.2.2) We could observed from the table 4.5 that the approach with highest average precision 0.7661 is the combination of (Trigram + Random forest learning algorithm + all data)

|  | Logistic regres-sion | Logistic regres-sion CV | SVM Linear SVC | SVM NuSVC | SVM SVC | Naive Bayes | Decision Tree | Random forest |
|---|---|---|---|---|---|---|---|---|
| Unigram | 0.7170 | 0.7151 | 0.6706 | 0.6439 | 0.7060 | 0.6016 | 0.7423 | 0.7352 |
| Bigram | 0.7063 | 0.7093 | 0.6914 | 0.6738 | 0.7103 | 0.5579 | 0.7671 | 0.7606 |
| Trigram | 0.6674 | 0.6593 | 0.6919 | 0.6538 | 0.6878 | 0.5492 | 0.7413 | 0.7436 |

Table 4.3: Average precision for all job positions of using work experience data

|  | Logistic regres-sion | Logistic regres-sion CV | SVM Linear SVC | SVM NuSVC | SVM SVC | Naive Bayes | Decision Tree | Random forest |
|---|---|---|---|---|---|---|---|---|
| Unigram | 0.5786 | 0.5745 | 0.5868 | 0.6330 | 0.5504 | 0.6672 | 0.5314 | 0.5086 |
| Bigram | 0.4993 | 0.5064 | 0.5289 | 0.5517 | 0.4972 | 0.7661 | 0.4386 | 0.4558 |
| Trigram | 0.4473 | 0.4696 | 0.3976 | 0.5056 | 0.4253 | 0.7492 | 0.4024 | 0.4222 |

Table 4.4: Average recall for all job positions of using work experience data

|  | Logistic regres-sion | Logistic regres-sion CV | SVM Linear SVC | SVM NuSVC | SVM SVC | Naive Bayes | Decision Tree | Random forest |
|---|---|---|---|---|---|---|---|---|
| Unigram | 0.6843 | 0.6796 | 0.6859 | 0.6540 | 0.6708 | 0.5534 | 0.7220 | 0.7181 |
| Bigram | 0.6936 | 0.6926 | 0.7036 | 0.6753 | 0.7107 | 0.5377 | 0.7398 | 0.7433 |
| Trigram | 0.6879 | 0.6952 | 0.7161 | 0.6757 | 0.6848 | 0.5571 | 0.7473 | 0.7661 |

Table 4.5: Average precision for all job positions of using education background data

|  | Logistic regres-sion | Logistic regres-sion CV | SVM Linear SVC | SVM NuSVC | SVM SVC | Naive Bayes | Decision Tree | Random forest |
|---|---|---|---|---|---|---|---|---|
| Unigram | 0.5786 | 0.5745 | 0.5868 | 0.6330 | 0.5504 | 0.6672 | 0.5314 | 0.5086 |
| Bigram | 0.5617 | 0.5672 | 0.5058 | 0.6030 | 0.4782 | 0.7352 | 0.5216 | 0.5305 |
| Trigram | 0.5585 | 0.5961 | 0.5113 | 0.6338 | 0.5497 | 0.6529 | 0.544 | 0.5285 |

Table 4.6: Average recall for all job positions of using education background data

4. Approaches using skills data: In this part, I use all skills data(i.e. including user skills data used in the chapter 3.2.2) We could observed from the table 4.3 that the approach with highest average precision 0.9015 is the combination of (Trigram + Random forest learning algorithm + all data)

|  | Logistic regres- sion | Logistic regres- sion CV | SVM Linear SVC | SVM NuSVC | SVM SVC | Naive Bayes | Decision Tree | Random forest |
|---|---|---|---|---|---|---|---|---|
| Unigram | 0.7661 | 0.7547 | 0.7300 | 0.7168 | 0.7436 | 0.6957 | 0.7688 | 0.7827 |
| Bigram | 0.8051 | 0.7626 | 0.7843 | 0.7512 | 0.7183 | 0.6639 | 0.8519 | 0.8566 |
| Trigram | 0.8168 | 0.8110 | 0.8621 | 0.7416 | 0.6794 | 0.6297 | 0.8970 | 0.9015 |

Table 4.7: Average precision for all job positions of using skills data

|  | Logistic regres- sion | Logistic regres- sion CV | SVM Linear SVC | SVM NuSVC | SVM SVC | Naive Bayes | Decision Tree | Random forest |
|---|---|---|---|---|---|---|---|---|
| Unigram | 0.6593 | 0.6982 | 0.6817 | 0.7069 | 0.7142 | 0.7236 | 0.5464 | 0.5965 |
| Bigram | 0.4654 | 0.6536 | 0.3968 | 0.5613 | 0.634 | 0.7610 | 0.4094 | 0.4145 |
| Trigram | 0.3097 | 0.3822 | 0.256 | 0.4686 | 0.4310 | 0.694 | 0.3090 | 0.3338 |

Table 4.8: Average recall for all job positions of using skills data

Analysis of the n-gram feature representation: Based on above tables, we could find that 83.3% of highlighted approaches are using Trigram for feature representation. Meanwhile, we could find that all highlighted approaches using Random forest as the learning algorithm for recommendation and the combination of (trigram + random forest learning algorithm + user skills data) outperforms other combinations with precision value of 0.9015. Thus, based on our analysis, the best content-based approach using N-grams as feature representation is Trigram + Random forest learning algorithm + user skills data. The approach with highest recall is the combination of bigram, naive bayes learning algorithm for all user data and the value is 0.774.

**Dimension reduction VS Non Dimension Reduction**

In our research we implemented the singular value decomposition in the N-grams feature representation technique in order to reduce the dimension of feature vectors. The dimension reduction have greatly reduced the training time of data. The table is shown in table4.9:

| | Logistic regression | Logistic regression CV | SVM Linear SVC | SVM NuSVC | SVM SVC | Naive Bayes | Decision Tree | Random forest |
|---|---|---|---|---|---|---|---|---|
| Reduced dimension | 0.004092 | 0.214267 | 0.044237 | 0.020129 | 0.016301 | 0.000796 | 0.013364 | 0.022350 |
| Non-reduced dimension | 0.549585 | 10.197601 | 0.526837 | 11.228063 | 11.486369 | 0.809088 | 1.394696 | 0.460641 |

Table 4.9: Training time of different learning algorithms with original data and reduced dimension data

We can observed that The best result of dimension reduction(highlighted red value) is Nave Bayes, reduced 99.9016% of original Training time.

**Document Embedding as Feature Representation in Content-based Approach**

Highest precision for different job positions using document embedding and different learning algorithms (Logistic regression, Logistic regression CV, SVM Linear SVC, SVM NuSVC, SVM SVC, Nave Bayes, Decision Tree, Random forest, ) are shown in table4.10: Then the average precision of job titles for every approaches are shown in table4.11:

Thus, based on our above tables, the approaches of document embedding + logistic regression algorithms have best performance with the highest average precision of all job titles.

| | Software engineer | Project manager | Research assistant | Process engineer | Consultant | Account manager |
|---|---|---|---|---|---|---|
| Learning algorithm | Logistic regression | Logistic regression | SVM Linear SVC | Logistic regression | SVM SVC | Logistic regression CV |
| Precision | 0.6463 | 0.5847 | 0.6680 | 0.6719 | 0.5621 | 0.5730 |

Table 4.10: Document embedding approaches with highest precision for every job position

| | Logistic regression | Logistic regression CV | SVM Linear SVC | SVM NuSVC | SVM SVC | Naive Bayes | Decision Tree | Random forest |
|---|---|---|---|---|---|---|---|---|
| Average Precision | 0.6165 | 0.6091 | 0.6096 | 0.6092 | 0.5636 | 0.5437 | 0.5301 | 0.5684 |

Table 4.11: Average precision for every learning algorithms with document embedding

| | Logistic regression | Logistic regression CV | SVM Linear SVC | SVM NuSVC | SVM SVC | Naive Bayes | Decision Tree | Random forest |
|---|---|---|---|---|---|---|---|---|
| Average recall | 0.5969 | 0.6045 | 0.5642 | 0.596 | 0.5525 | 0.5670 | 0.5418 | 0.4601 |

Table 4.12: Average recall for every learning algorithms with document embedding

### Word Embedding as Feature Representation in Content-based Approach

The result of word embedding with different learning algorithms (Logistic regression, Logistic regression CV, SVM Linear SVC, SVM NuSVC, SVM SVC, Nave Bayes, Decision Tree, Random forest, ) is shown in table 4.13. Then we calculate average precision of job positions for every approaches, the result is shown in table 4.15

We could see from table 4.13 that highest precision of recommendation for job title 'software engineer', 'research assistant' and 'process engineer' is much higher than pre- cision of job title 'consultant', 'project manager' and 'account manager'. Based on the

|  | Software engi-neer | Project man-ager | Research assis-tant | Process engi-neer | Consultant | Account man-ager |
|---|---|---|---|---|---|---|
| Learning algo-rithm | Random forest | Logistic regres-sion | SVM NuSVC | Logistic regres-sion CV | Logistic regression | SVM Linear SVC |
| Precision | 0.8577 | 0.6698 | 0.8300 | 0.8097 | 0.6009 | 0.6930 |

Table 4.13: Word embedding approaches with highest precision for every job position

|  | Logistic regres-sion | Logistic regres-sion CV | SVM Linear SVC | SVM NuSVC | SVM SVC | Naive Bayes | Decision Tree | Random forest |
|---|---|---|---|---|---|---|---|---|
| Average Preci-sion | 0.74 | 0.7328 | 0.7275 | 0.7242 | 0.6729 | 0.6555 | 0.6563 | 0.7184 |

Table 4.14: Average precision for every learning algorithms with word  embedding

|  | Logistic regres-sion | Logistic regres-sion CV | SVM Linear SVC | SVM NuSVC | SVM SVC | Naive Bayes | Decision Tree | Random forest |
|---|---|---|---|---|---|---|---|---|
| Average recall | 0.656 | 0.6592 | 0.6357 | 0.6406 | 0.7398 | 0.7113 | 0.5413 | 0.5284 |

Table 4.15: Average recall for every learning algorithms with word embedding

table 4.15, the word embedding approach + logistic regression algorithms outperforms other word embedding approach with the average precision of 0.74.

## 4.2.2   Evaluation for Collaborative Filtering Approaches

The result of the collaborative filtering approach with learning algorithms (SVDplus-plus, ASVDplusplus ) is shown in table 4.16:

Based on our results, we can observed that ASVDplusplus learning algorithms has better performance than SVDplusplus learning algorithm because ASVDplusplus has higher precision. Then we calculated the average precision of the two collaborative approaches and get results. The average precision of ASVDplusplus is 0.7173, which is higher than the average precision 0.6526 of SVDplusplus learning algorithm. Then value of average recall of two algorithms are 0.613 (ASVDplusplus) and 0.5092(SVD-plusplus)

|              | Software engineer | Project manager | Research assistant | Process engineer | Consultant | Account manager |
|--------------|-------------------|-----------------|--------------------|------------------|------------|-----------------|
| ASVDplusplus | 0.6926            | 0.7131          | 0.6978             | 0.7037           | 0.7309     | 0.7661          |
| SVDplusplus  | 0.6696            | 0.6586          | 0.6228             | 0.6106           | 0.6875     | 0.6667          |

Table 4.16: Collaborative filtering approaches with highest precision for every job position

## 4.3   Discussion of Results

We have calculated the average precision of different job titles using a content-based approach with different feature representations, and using a collaborative filtering approach. We will discuss the results based on the metrics used.

The approach with highest precision is the content-based approach combining trigrams with the random forest learning algorithm. This approach outperforms other content-based and collaborative filtering approaches with a precision of 0.9015, which means 90% of the predicting positive is correct. Other approach's average precision are between 0.7 to 0.9. The approach with the highest recall is the content-based approach combining bigrams with naive bayes and all user data. The recall value is 0.774, which means 77.4% of positive samples are predicted positive in average for all job titles. We also find that this approach is still the best performing in terms of recall when we tried all different parts of the data.

Meanwhile, by implementing singular value decomposition for reducing the dimensionality of feature vectors, we have found that training time is greatly reduced, which is significant for online recommendation because users usually need real-time recommendation in an online environment (i.e. users on job websites will receive real-time recommendation of different jobs based on operations like updating their user profile and viewing job history).

# Chapter 5

# Conclusion and Future Work

In this chapter, we will conclude our findings of different approaches and the perfor-
mance of our recommender engine. At the end of this chapter, we will discuss the
futures works of our project and share some thoughts about that.

### 5.0.1   Conclusion of the Recommender System

The recommender engine developed by this research is designed to find the best
approach for recommending the job title by combining machine learning and language
models. This engine has been successfully implemented.

### 5.0.2   Conclusion of Results

The results presented in Chapter 4 illustrated that the content-based approach of
combining trigrams with the random forest learning algorithm and skills has the
highest average precision for all job titles in talent recommendation. The content-based
approach of combining bigrams with the naive bayes learning algorithm and all user
data has the highest average recall for all job titles. Thus from the perspective of
accurate prediction, a combination of the trigrams, random forest learning algorithm
and skills could support the most effective talent recommendation.  From  the
perspective of high coverage of the correct prediction, the combination of bigrams,
naive bayes learning algorithm and all user data is the most suitable approach for
talent recommendation. The singular value decomposition technique for dimension
reduction has shown great effectiveness in reducing the training time of the feature data
by 99%, which would greatly improve the efficiency of talent recommendation.

### 5.0.3   Future Works

We have identified the approaches which have the best precision@n and recall@n by
testing different machine learning algorithms and different language models.  In the
future,  we will focus on the following aspects to improve our recommender engine:

1.*Using practical ratios of data*: the training set and testing set in current project is small, and the ratio of the data (50% positive 50% negative) is idealized, in the following steps of the optimization, we will test with real ratios (such as 60 relevant and 1520 non-relevant users) and change the ratio to more accurately mirror the real world situation faced by talent search.

2. *Consider implementing more techniques based on state-of-the-art research.* Other state-of-the-art research such as knowledge graph embedding could also be considered for talent recommendation (i.e. knowledge graph embedding). We could also consider building a hybrid recommender engine to maintain the precision and quality of recommendation (i.e. for top 200 recommendation, the top 30 result can be selected from approaches with high precision, while the rest of the recommendation can use the results from high recall value).

# Appendix

| software engineer | Unigram | Bigram | Trigram |
|---|---|---|---|
| Logistic regression | 0.849837207 | 0.806616508 | 0.770752284 |
| Logistic regression CV | 0.83547896 | 0.780248088 | 0.731117624 |
| SVM Linear SVC | 0.853768979 | 0.871361402 | 0.857921171 |
| SVM NuSVC | 0.717154746 | 0.806425287 | 0.795387517 |
| SVM SVC | 0.782923429 | 0.80559273 | 0.809540369 |
| Nave Bayes | 0.70934625 | 0.609165442 | 0.537402642 |
| Decision Tree | 0.853003826 | 0.85602009 | 0.870084196 |
| Random forest | 0.840626731 | 0.854964393 | 0.888328989 |

| project manager | Unigram | Bigram | Trigram |
|---|---|---|---|
| Logistic regression | 0.647392108 | 0.682261384 | 0.613923859 |
| Logistic regression CV | 0.681774039 | 0.654091793 | 0.57925159 |
| SVM Linear SVC | 0.615012127 | 0.695389595 | 0.476258521 |
| SVM NuSVC | 0.618691705 | 0.650219942 | 0.599921248 |
| SVM SVC | 0.614239442 | 0.672558586 | 0.624614068 |
| Nave Bayes | 0.566678481 | 0.530981183 | 0.519475202 |
| Decision Tree | 0.663664304 | 0.694715146 | 0.736639657 |
| Random forest | 0.666557866 | 0.72747995 | 0.702783479 |

The precision of the different approaches for

| research assistant | Unigram | Bigram | Trigram |
|---|---|---|---|
| Logistic regression | 0.872883391 | 0.871789863 | 0.868190284 |
| Logistic regression CV | 0.852958763 | 0.837562048 | 0.841421171 |
| SVM Linear SVC | 0.89858578 | 0.911761662 | 0.956708408 |
| SVM NuSVC | 0.68061625 | 0.845216286 | 0.854786417 |
| SVM SVC | 0.787943963 | 0.857103711 | 0.856810509 |
| Nae Bayes | 0.689872503 | 0.603165372 | 0.557270601 |
| Decision Tree | 0.793007922 | 0.813914291 | 0.824471249 |
| Random forest | 0.856757629 | 0.875588554 | 0.861596308 |

| process engineer | Unigram | Bigram | Trigram |
|---|---|---|---|
| Logistic regression | 0.815324603 | 0.735207644 | 0.706229407 |
| Logistic regression CV | 0.803093085 | 0.741061878 | 0.665343829 |
| SVM Linear SVC | 0.814664681 | 0.726335239 | 0.843175477 |
| SVM NuSVC | 0.768170731 | 0.774029548 | 0.716029723 |
| SVM SVC | 0.803498511 | 0.761145208 | 0.720354926 |
| Nae Bayes | 0.72858614 | 0.601990764 | 0.583597028 |
| Decision Tree | 0.771395682 | 0.810583745 | 0.818481187 |
| Random forest | 0.796954286 | 0.789746586 | 0.849515507 |

| consultant | Unigram | Bigram | Trigram |
|---|---|---|---|
| Logistic regression | 0.58674806 | 0.612450783 | 0.651725443 |
| Logistic regression CV | 0.57972959 | 0.607656041 | 0.614986594 |
| SVM Linear SVC | 0.599837624 | 0.631315789 | 0.83 |
| SVM NuSVC | 0.557083065 | 0.590817531 | 0.635084372 |
| SVM SVC | 0.564665983 | 0.618522693 | 0.66236589 |
| Nae Bayes | 0.516111887 | 0.516496723 | 0.506864346 |
| Decision Tree | 0.604761077 | 0.678704275 | 0.696484792 |
| Random forest | 0.621355423 | 0.674134512 | 0.736430047 |

| account manager | Unigram | Bigram | Trigram |
|---|---|---|---|
| Logistic regression | 0.685271927 | 0.671397091 | 0.638191279 |
| Logistic regression CV | 0.715928283 | 0.644638195 | 0.618668869 |
| SVM Linear SVC | 0.630880097 | 0.678552107 | 0.726289513 |
| SVM NuSVC | 0.694804714 | 0.646574091 | 0.630672766 |
| SVM SVC | 0.741223665 | 0.61691456 | 0.583416647 |
| Nae Bayes | 0.655963619 | 0.643804922 | 0.593822572 |
| Decision Tree | 0.714099684 | 0.723243415 | 0.719721766 |
| Random forest | 0.682095506 | 0.67166967 | 0.746477241 |

Table 1:

| | Unigram | | | | Bigram | | | | Trigram | | | |
|---|---|---|---|---|---|---|---|---|---|---|---|---|
| | All | Work experience | Education background | Skills | All | Work experience | Education background | Skills | All | Work experience | Education background | Skills |
| Logistic regression | 0.849837207 | 0.773428678 | 0.80244605 | 0.919109499 | 0.806616508 | 0.801883665 | 0.836343813 | 0.930033359 | 0.770752284 | 0.746262525 | 0.842291536 | 0.938483019 |
| Logistic regression CV | 0.83547896 | 0.775168281 | 0.789305356 | 0.895318169 | 0.780248088 | 0.806591044 | 0.818029677 | 0.863133479 | 0.731117624 | 0.739858137 | 0.817836696 | 0.934047452 |
| SVM Linear SVC | 0.857568979 | 0.724778606 | 0.7782547 | 0.904504752 | 0.871361402 | 0.782589 | 0.840853714 | 0.950988509 | 0.857921711 | 0.781917721 | 0.870327821 | 0.966753708 |
| SVM NuSVC | 0.717154746 | 0.681137699 | 0.755886446 | 0.835547897 | 0.806425287 | 0.74587029 | 0.82505971 | 0.8961629 | 0.79538751 | 0.748769888 | 0.845425752 | 0.925267822 |
| SVM SVC | 0.782923429 | 0.748791148 | 0.762546371 | 0.912164113 | 0.80659273 | 0.786582339 | 0.841611309 | 0.822320899 | 0.80954037 | 0.763002925 | 0.777975151 | 0.599458623 |
| Nae Bayes | 0.70934625 | 0.588931371 | 0.574718949 | 0.859067023 | 0.691654842 | 0.540999118 | 0.541499368 | 0.807006857 | 0.537402642 | 0.505695152 | 0.564821198 | 0.788998875 |
| Decision Tree | 0.853003826 | 0.821017404 | 0.829018812 | 0.922589029 | 0.85902009 | 0.856609613 | 0.876626501 | 0.947199981 | 0.837008196 | 0.882078419 | 0.913023399 | 0.961018854 |
| Random forest | 0.840626731 | 0.779755162 | 0.815054245 | 0.930854576 | 0.854964393 | 0.865117308 | 0.865116049 | 0.958810604 | 0.888328989 | 0.849674812 | 0.907921443 | 0.959489417 |

| Project manager: | Unigram | | | | Bigram | | | | Trigram | | | |
|---|---|---|---|---|---|---|---|---|---|---|---|---|
| | All | Work experience | Education background | Skills | All | Work experience | Education background | Skills | All | Work experience | Education background | Skills |
| Logistic regression | 0.647392108 | 0.642178802 | 0.568277085 | 0.696833732 | 0.682261384 | 0.63169524 | 0.596835758 | 0.80154806 | 0.603923859 | 0.592640879 | 0.593777301 | 0.74359046 |
| Logistic regression CV | 0.681734039 | 0.651010379 | 0.57083688 | 0.59974597 | 0.654091793 | 0.636694963 | 0.584296142 | 0.76709352 | 0.579251159 | 0.587644813 | 0.61095445 | 0.749062927 |
| SVM Linear SVC | 0.615012127 | 0.588467265 | 0.591324712 | 0.604897081 | 0.695589595 | 0.583415344 | 0.616571072 | 0.572177365 | 0.599972126 | 0.599911322 | 0.506646195 | 0.659683392 |
| SVM NuSVC | 0.618691705 | 0.58484995 | 0.564668382 | 0.66755681 | 0.650219942 | 0.58615719 | 0.538773499 | 0.721135663 | 0.599021459 | 0.589311322 | 0.585207018 | 0.659894388 |
| SVM SVC | 0.614239442 | 0.620706737 | 0.57245035 | 0.686223728 | 0.672258586 | 0.627947304 | 0.587763837 | 0.749183536 | 0.624614068 | 0.613383328 | 0.532674855 | 0.783070637 |
| Nae Bayes | 0.56667846 | 0.538998612 | 0.549818408 | 0.572320643 | 0.50999183 | 0.52170288 | 0.51480193 | 0.529834416 | 0.519475202 | 0.506964104 | 0.566533991 | 0.506354739 |
| Decision Tree | 0.663664304 | 0.665845238 | 0.634655293 | 0.682759992 | 0.694715146 | 0.677795449 | 0.66515232 | 0.656021756 | 0.736639657 | 0.674895352 | 0.729520731 | 0.890533866 |
| Random forest | 0.666557866 | 0.639697497 | 0.614406122 | 0.717914645 | 0.72747995 | 0.661731861 | 0.665152322 | 0.630547473 | 0.702783479 | 0.661139781 | 0.735657612 | 0.902690945 |

| Research assistant: | Unigram | | | | Bigram | | | | Trigram | | | |
|---|---|---|---|---|---|---|---|---|---|---|---|---|
| | All | Work experience | Education background | Skills | All | Work experience | Education background | Skills | All | Work experience | Education background | Skills |
| Logistic regression | 0.872883391 | 0.830419127 | 0.804191127 | 0.850805134 | 0.871789863 | 0.805242739 | 0.866365611 | 0.823827025 | 0.868190284 | 0.772047321 | 0.834616913 | 0.900869963 |
| Logistic regression CV | 0.852958763 | 0.822502612 | 0.781029288 | 0.796448197 | 0.835766344 | 0.830611574 | 0.828815444 | 0.784488743 | 0.841421171 | 0.773325738 | 0.811270841 | 0.848140711 |
| SVM Linear SVC | 0.899858578 | 0.736225737 | 0.768141049 | 0.867120379 | 0.919176162 | 0.755322137 | 0.830054305 | 0.832757558 | 0.959678408 | 0.719144544 | 0.846980587 | 0.976190476 |
| SVM NuSVC | 0.68061625 | 0.710038256 | 0.742559452 | 0.690507743 | 0.845216286 | 0.750075143 | 0.830765672 | 0.794171742 | 0.854786417 | 0.702839753 | 0.831138055 | 0.853186891 |
| SVM SVC | 0.787943963 | 0.871668296 | 0.816816433 | 0.685682278 | 0.857103711 | 0.844117453 | 0.926624772 | 0.627524378 | 0.856810599 | 0.808729753 | 0.910218283 | 0.625472506 |
| Nae Bayes | 0.689872503 | 0.782996086 | 0.555147431 | 0.775616311 | 0.603165372 | 0.658430633 | 0.522509693 | 0.777392875 | 0.53772001 | 0.660808283 | 0.516567828 | 0.75880394 |
| Decision Tree | 0.793007922 | 0.82101629 | 0.79682017 | 0.819955166 | 0.813914291 | 0.822589594 | 0.832540046 | 0.816271112 | 0.824241249 | 0.721391949 | 0.718461166 | 0.88962227 |
| Random forest | 0.856757629 | 0.829017819 | 0.822450332 | 0.856367175 | 0.875588554 | 0.820051763 | 0.851847878 | 0.859950912 | 0.861596308 | 0.791184179 | 0.83809788 | 0.926943442 |

| Process engineer: | Unigram | | | | Bigram | | | | Trigram | | | |
|---|---|---|---|---|---|---|---|---|---|---|---|---|
| | All | Work experience | Education background | Skills | All | Work experience | Education background | Skills | All | Work experience | Education background | Skills |
| Logistic regression | 0.837524403 | 0.742908458 | 0.766449546 | 0.838681363 | 0.735207644 | 0.734056479 | 0.780146146 | 0.867679444 | 0.706229407 | 0.707387409 | 0.783789412 | 0.897106833 |
| Logistic regression CV | 0.803093083 | 0.740737608 | 0.770306492 | 0.817469231 | 0.714061878 | 0.739427003 | 0.786186017 | 0.855268169 | 0.663543829 | 0.686617069 | 0.773342012 | 0.874607266 |
| SVM Linear SVC | 0.814664681 | 0.697898164 | 0.738190688 | 0.791321935 | 0.768563663 | 0.72635239 | 0.742061138 | 0.822752101 | 0.726457644 | 0.843178477 | 0.759401877 | 0.860102452 |
| SVM NuSVC | 0.576817031 | 0.670084 | 0.758511129 | 0.637630983 | 0.774029548 | 0.717036805 | 0.782227206 | 0.754825642 | 0.716629723 | 0.729336757 | 0.814042234 | 0.718443743 |
| SVM SVC | 0.803498511 | 0.708174807 | 0.725234135 | 0.8738284 | 0.761145208 | 0.727666469 | 0.80370321 | 0.753860261 | 0.720354926 | 0.745939901 | 0.813991074 | 0.654719836 |
| Nae Bayes | 0.72858602 | 0.598477724 | 0.567015383 | 0.796995472 | 0.600990764 | 0.538599956 | 0.537342008 | 0.516631908 | 0.589708782 | 0.620219482 | 0.520219482 | 0.71271248 |
| Decision Tree | 0.771195682 | 0.78221564 | 0.816868657 | 0.83262182 | 0.81058745 | 0.800253405 | 0.649864677 | 0.934440859 | 0.656746773 | 0.855887967 | 0.947113723 | |
| Random forest | 0.769954286 | 0.773926566 | 0.804520644 | 0.857910274 | 0.789746586 | 0.805464023 | 0.852461861 | 0.929174923 | 0.849515507 | 0.814247185 | 0.893159016 | 0.947325694 |

| Consultant: | Unigram | | | | Bigram | | | | Trigram | | | |
|---|---|---|---|---|---|---|---|---|---|---|---|---|
| | All | Work experience | Education background | Skills | All | Work experience | Education background | Skills | All | Work experience | Education background | Skills |
| Logistic regression | 0.58674806 | 0.601821999 | 0.563125479 | 0.601246741 | 0.612450783 | 0.581999605 | 0.529071285 | 0.639354567 | 0.651725443 | 0.577914228 | 0.552860628 | 0.604126984 |
| Logistic regression CV | 0.57972959 | 0.59325642 | 0.569161173 | 0.603669252 | 0.607656041 | 0.585058271 | 0.544675554 | 0.592579769 | 0.614496594 | 0.579875911 | 0.562595289 | 0.77591154 |
| SVM Linear SVC | 0.599837624 | 0.609687562 | 0.593711267 | 0.610905345 | 0.643135571 | 0.554655554 | 0.560727481 | 0.66474138 | 0.712628913 | 0.643004065 | 0.519848766 | 0.701818182 |
| SVM NuSVC | 0.557083065 | 0.554020621 | 0.532387642 | 0.576678588 | 0.590873807 | 0.583785439 | 0.593951654 | 0.606069075 | 0.635084372 | 0.558609698 | 0.543435503 | 0.606565336 |
| SVM SVC | 0.564665983 | 0.572243573 | 0.559583623 | 0.580177851 | 0.614852693 | 0.601742369 | 0.553119461 | 0.619522799 | 0.662365589 | 0.599027985 | 0.559975522 | 0.624777328 |
| Nae Bayes | 0.516611887 | 0.531102634 | 0.50303808 | 0.51451136 | 0.516496723 | 0.494970972 | 0.509483021 | 0.512657256 | 0.50684488 | 0.492807146 | 0.505542099 | 0.501643867 |
| Decision Tree | 0.604761077 | 0.621479026 | 0.610043562 | 0.652223568 | 0.67702275 | 0.678235662 | 0.586272927 | 0.743539383 | 0.694647942 | 0.60922382 | 0.611946705 | 0.810830056 |
| Random forest | 0.621355423 | 0.645750694 | 0.616000234 | 0.639348472 | 0.674134512 | 0.684247949 | 0.596324877 | 0.724090785 | 0.736430047 | 0.654688512 | 0.650863511 | 0.826900123 |

| Account manager: | Unigram | | | | Bigram | | | | Trigram | | | |
|---|---|---|---|---|---|---|---|---|---|---|---|---|
| | All | Work experience | Education background | Skills | All | Work experience | Education background | Skills | All | Work experience | Education background | Skills |
| Logistic regression | 0.685721927 | 0.723199858 | 0.601571231 | 0.690196058 | 0.671397091 | 0.683320002 | 0.55320696 | 0.568370912 | 0.643912279 | 0.650863596 | 0.520633746 | 0.816173026 |
| Logistic regression CV | 0.717928283 | 0.70823692 | 0.597031354 | 0.705916531 | 0.644638195 | 0.68218979 | 0.591749997 | 0.707563459 | 0.636846869 | 0.594559501 | 0.562595289 | 0.779911454 |
| SVM Linear SVC | 0.630880097 | 0.66663587 | 0.592915002 | 0.618515099 | 0.678552107 | 0.631570967 | 0.761709671 | 0.726289513 | 0.643004065 | 0.519848766 | 0.869138439 | |
| SVM NuSVC | 0.694804714 | 0.663411078 | 0.58808484 | 0.693293792 | 0.644574091 | 0.6069168161 | 0.535834508 | 0.729064096 | 0.604627562 | 0.59417378 | 0.51899784 | 0.786843127 |
| SVM SVC | 0.741223665 | 0.708545598 | 0.626237883 | 0.723606637 | 0.610991456 | 0.673954877 | 0.5353537739 | 0.741189059 | 0.583416647 | 0.596880536 | 0.518348259 | 0.789495431 |
| Nae Bayes | 0.655963619 | 0.645697275 | 0.565401878 | 0.657927039 | 0.54558904 | 0.600712759 | 0.53889916 | 0.993822572 | 0.602805745 | 0.660916008 | 0.509880457 | |
| Decision Tree | 0.714099684 | 0.742441769 | 0.644749789 | 0.703195206 | 0.723243415 | 0.67131739 | 0.632257706 | 0.83428058 | 0.519721766 | 0.626369537 | 0.57848807 | 0.883319293 |
| Random forest | 0.682095506 | 0.743135306 | 0.636580218 | 0.694137133 | 0.671166967 | 0.728929751 | 0.629105675 | 0.837655813 | 0.746477241 | 0.69094522 | 0.571368271 | 0.84597139 |

# Precision of different job titles (reduce dimension to 50) with different combination of user data

| | Unigram | | | | Bigram | | | | Trigram | | | |
|---|---|---|---|---|---|---|---|---|---|---|---|---|
| | All | Work experience | Education background | Skills | All | Work experience | Education background | Skills | All | Work experience | Education background | Skills |
| Logistic regression | 0.798747525 | 0.749873729 | 0.786680096 | 0.884126523 | 0.710565127 | 0.710565127 | 0.925739085 | 0.610890386 | 0.707239001 | 0.698792909 | 0.813829282 | 0.767141826 |
| Logistic regression CV | 0.794443219 | 0.763168853 | 0.80387804 | 0.881008863 | 0.720173679 | 0.720173679 | 0.838014701 | 0.802151879 | 0.751401391 | 0.757953044 | 0.833837724 | 0.749773552 |
| SVM Linear SVC | 0.878650588 | 0.660635567 | 0.592915002 | 0.619497679 | 0.91907739 | 0.812411325 | 0.721214325 | 0.909508042 | 0.84570828 | 0.728811969 | 0.82166289 | 0.84610338 |
| SVM NuSVC | 0.834973468 | 0.709990552 | 0.818377796 | 0.947217916 | 0.833714134 | 0.837714134 | 0.916526279 | 0.942812743 | 0.903464408 | 0.821631626 | 0.896529422 | 0.947495057 |
| SVM SVC | 0.83123108 | 0.688458814 | 0.769378243 | 0.939273967 | 0.76068637 | 0.61691456 | 0.627940845 | 0.512299231 | 0.617641567 | 0.676091891 | 0.777570047 | 0.500215501 |
| Nae Bayes | 0.5 | 0.5 | 0.5 | 0.5 | 0.501833054 | 0.501833054 | 0.508040802 | 0.5 | 0.499979716 | 0.5 | 0.497199062 | 0.500200401 |
| Decision Tree | 0.818225396 | 0.56981572 | 0.579378263 | 0.777374849 | 0.552713108 | 0.552713108 | 0.49349838 | 0.699187583 | 0.936698462 | 0.591116376 | 0.646108574 | 0.683379753 |
| Random forest | 0.689412539 | 0.670620228 | 0.654556293 | 0.824096156 | 0.612271701 | 0.612289701 | 0.644378268 | 0.741965527 | 0.703906288 | 0.578685856 | 0.627114165 | 0.806581524 |

| Project manager: | Unigram | | | | Bigram | | | | Trigram | | | |
|---|---|---|---|---|---|---|---|---|---|---|---|---|
| | All | Work experience | Education background | Skills | All | Work experience | Education background | Skills | All | Work experience | Education background | Skills |
| Logistic regression | 0.676038738 | 0.644299522 | 0.583235691 | 0.716255283 | 0.608250721 | 0.608250721 | 0.615835392 | 0.736155797 | 0.641463794 | 0.60848478 | 0.70613138 | 0.689836431 |
| Logistic regression CV | 0.657123467 | 0.635393782 | 0.618514012 | 0.712453635 | 0.650371674 | 0.650371674 | 0.601307462 | 0.729196148 | 0.636204605 | 0.629549324 | 0.726811969 | 0.680044584 |
| SVM Linear SVC | 0.733639609 | 0.571765248 | 0.727109002 | 0.804672703 | 0.717107763 | 0.717107763 | 0.750071283 | 0.809458381 | 0.738244268 | 0.729120104 | 0.566754852 | 0.840483363 |
| SVM SVC | 0.654725032 | 0.641496938 | 0.556359771 | 0.709875611 | 0.642717 | 0.64954742 | 0.559677344 | 0.726390186 | 0.631415992 | 0.614749995 | 0.558393816 | 0.642936318 |
| Nae Bayes | 0.499193548 | 0.5 | 0.5 | 0.5 | 0.5 | 0.5 | 0.499799599 | 0.5 | 0.500684 | 0.5 | 0.501030032 | 0.5 |
| Decision Tree | 0.549975318 | 0.531790262 | 0.512942223 | 0.587663863 | 0.5225697 | 0.5225697 | 0.510949077 | 0.647488889 | 0.549925616 | 0.589006166 | 0.509924659 | 0.569233054 |
| Random forest | 0.600907022 | 0.562243325 | 0.575991156 | 0.612163169 | 0.580353972 | 0.580351918 | 0.634918646 | 0.533282642 | 0.540294213 | 0.561549219 | 0.609508471 | |

**Research assistant:**

| | Unigram | | | | Bigram | | | | Trigram | | | |
|---|---|---|---|---|---|---|---|---|---|---|---|---|
| | All | Work experience | Education background | Skills | All | Work experience | Education background | Skills | All | Work experience | Education background | Skills |
| Logistic regression | 0.797360071 | 0.786460197 | 0.801119913 | 0.819436883 | 0.787585736 | 0.787585736 | 0.797098554 | 0.669171008 | 0.79785241 | 0.778809073 | 0.837736867 | 0.622181405 |
| Logistic regression CV | 0.79899842 | 0.816601646 | 0.786724437 | 0.814821982 | 0.75211419 | 0.75211419 | 0.789886463 | 0.702257592 | 0.781166639 | 0.795041583 | 0.826793064 | 0.650797437 |
| SVM Linear SVC | 0.804118546 | 0.786791695 | 0.824492297 | 0.796299573 | 0.803970987 | 0.803970987 | 0.796570563 | 0.744477584 | 0.804477449 | 0.762428681 | 0.834292719 | 0.681558137 |
| SVM NuSVC | 0.821475265 | 0.784931389 | 0.791589252 | 0.861525014 | 0.737870136 | 0.737870136 | 0.75968112 | 0.814128924 | 0.796049613 | 0.802847111 | 0.81047592 | 0.672909299 |
| SVM SVC | 0.790363968 | 0.715143966 | 0.827759701 | 0.826993495 | 0.764239359 | 0.764239359 | 0.810079604 | 0.519860995 | 0.77211789 | 0.707122113 | 0.82258204 | 0.518379172 |
| Naive Bayes | 0.528831857 | 0.50020284 | 0.5 | 0.5 | 0.5 | 0.5 | 0.499996748 | 0.500755668 | 0.499595155 | 0.500526192 | 0.498768671 | 0.499598394 |
| Decision Tree | 0.563327231 | 0.514247397 | 0.536882646 | 0.646445824 | 0.59852365 | 0.59852365 | 0.568567508 | 0.486086108 | 0.560054011 | 0.560351154 | 0.499724211 | 0.558890322 |
| Random forest | 0.613170301 | 0.583037422 | 0.542125929 | 0.773060882 | 0.630576975 | 0.630576975 | 0.476214134 | 0.632923352 | 0.710910168 | 0.527347389 | 0.62493183 | 0.608538196 |

**Process engineer:**

| | Unigram | | | | Bigram | | | | Trigram | | | |
|---|---|---|---|---|---|---|---|---|---|---|---|---|
| | All | Work experience | Education background | Skills | All | Work experience | Education background | Skills | All | Work experience | Education background | Skills |
| Logistic regression | 0.794683568 | 0.737078947 | 0.735399577 | 0.855639295 | 0.689059227 | 0.689059227 | 0.761162357 | 0.817097586 | 0.691029948 | 0.703299407 | 0.74393667 | 0.754777488 |
| Logistic regression CV | 0.796697023 | 0.742033303 | 0.779934425 | 0.85667552 | 0.699040109 | 0.699040109 | 0.800185568 | 0.777222538 | 0.721851268 | 0.714731571 | 0.755063841 | 0.749617972 |
| SVM Linear SVC | 0.88918843 | 0.8358455 | 0.89993052 | 0.889700697 | 0.852160719 | 0.852160719 | 0.920782964 | 0.823793313 | 0.831905061 | 0.865326723 | 0.865552986 | 0.809580931 |
| SVM NuSVC | 0.832214116 | 0.715438614 | 0.856667182 | 0.858981409 | 0.849564724 | 0.849564724 | 0.863383527 | 0.939356086 | 0.899497885 | 0.914224003 | 0.820548081 | 0.926625266 |
| SVM SVC | 0.846087253 | 0.704287988 | 0.753026125 | 0.829750764 | 0.728302014 | 0.728302014 | 0.737330874 | 0.72975311 | 0.706315801 | 0.686984154 | 0.5 | 0.528000718 |
| Naive Bayes | 0.5 | 0.5 | 0.5 | 0.5 | 0.5 | 0.5 | 0.500401608 | 0.501538462 | 0.5 | 0.5 | 0.5 | 0.5 |
| Decision Tree | 0.627632795 | 0.594363283 | 0.633658581 | 0.556076714 | 0.575073264 | 0.575073264 | 0.489014134 | 0.694040538 | 0.643743971 | 0.563779484 | 0.581378206 | 0.593299306 |
| Random forest | 0.658925877 | 0.635097351 | 0.611337148 | 0.662859896 | 0.625423289 | 0.625423289 | 0.569758086 | 0.71424445 | 0.606238218 | 0.649120811 | 0.529462577 | 0.689479799 |

**Consultant:**

| | Unigram | | | | Bigram | | | | Trigram | | | |
|---|---|---|---|---|---|---|---|---|---|---|---|---|
| | All | Work experience | Education background | Skills | All | Work experience | Education background | Skills | All | Work experience | Education background | Skills |
| Logistic regression | 0.575387289 | 0.570119183 | 0.566875963 | 0.600178812 | 0.571137069 | 0.571137069 | 0.603668007 | 0.635119653 | 0.593550983 | 0.580683405 | 0.629177021 | 0.646432315 |
| Logistic regression CV | 0.578261006 | 0.58208796 | 0.595965532 | 0.600678741 | 0.63231299 | 0.63231299 | 0.593269794 | 0.622707343 | 0.591982826 | 0.596364873 | 0.613095374 | 0.640278782 |
| SVM Linear SVC | 0.734313274 | 0.685025977 | 0.671851688 | 0.686616624 | 0.731312514 | 0.731312514 | 0.68334277 | 0.710943899 | 0.674657593 | 0.670177408 | 0.704396257 | 0.710133788 |
| SVM NuSVC | 0.617801589 | 0.579823725 | 0.643554809 | 0.619497614 | 0.606706006 | 0.606706006 | 0.644809579 | 0.726690536 | 0.637004403 | 0.642145582 | 0.652980173 | 0.678016793 |
| SVM SVC | 0.612607313 | 0.566878325 | 0.526535561 | 0.626332608 | 0.614494585 | 0.614494585 | 0.59505923 | 0.65009894 | 0.641268773 | 0.614372248 | 0.591307987 | 0.661101666 |
| Naive Bayes | 0.499799599 | 0.499598394 | 0.499561671 | 0.499598458 | 0.499598394 | 0.499598394 | 0.500801603 | 0.5 | 0.519193535 | 0.5 | 0.498166919 | 0.497654584 |
| Decision Tree | 0.569217685 | 0.5185051 | 0.515675395 | 0.573356608 | 0.52961322 | 0.52961322 | 0.53383676 | 0.52049665 | 0.514241055 | 0.525377414 | 0.519768242 | 0.546219703 |
| Random forest | 0.5227667 | 0.51524192 | 0.541195647 | 0.521006562 | 0.493792676 | 0.493792676 | 0.534217296 | 0.482598103 | 0.518460104 | 0.50529053 | 0.530181078 | 0.506763043 |

**Account manager:**

| | Unigram | | | | Bigram | | | | Trigram | | | |
|---|---|---|---|---|---|---|---|---|---|---|---|---|
| | All | Work experience | Education background | Skills | All | Work experience | Education background | Skills | All | Work experience | Education background | Skills |
| Logistic regression | 0.699008619 | 0.702957659 | 0.626574162 | 0.731650242 | 0.640227368 | 0.640227368 | 0.658958506 | 0.754602466 | 0.610371462 | 0.605352107 | 0.651420306 | 0.663459435 |
| Logistic regression CV | 0.698518311 | 0.698672273 | 0.626765742 | 0.738300224 | 0.656703013 | 0.656703013 | 0.637028503 | 0.753690762 | 0.639952288 | 0.616224411 | 0.677602255 | 0.667105668 |
| SVM Linear SVC | 0.754043006 | 0.750099571 | 0.694547657 | 0.808740598 | 0.705667131 | 0.705667131 | 0.731348132 | 0.61792334 | 0.721243159 | 0.717983593 | 0.732302869 | 0.749410662 |
| SVM NuSVC | 0.747662396 | 0.743929878 | 0.666137415 | 0.803754064 | 0.610553584 | 0.610553584 | 0.657746248 | 0.903038031 | 0.781914824 | 0.777827726 | 0.802882086 | 0.689059352 |
| SVM SVC | 0.766731995 | 0.665401109 | 0.639846567 | 0.759430034 | 0.656112477 | 0.656112477 | 0.587040309 | 0.796506203 | 0.547361311 | 0.531057426 | 0.605189724 | 0.543250664 |
| Naive Bayes | 0.499799599 | 0.5 | 0.5 | 0.499596378 | 0.5 | 0.5 | 0.503124314 | 0.499597993 | 0.499796999 | 0.499799599 | 0.501375055 | 0.5 |
| Decision Tree | 0.553648642 | 0.527679245 | 0.542462631 | 0.592042921 | 0.541851305 | 0.541851305 | 0.550228986 | 0.56183495 | 0.569370078 | 0.558005703 | 0.604153946 | 0.600134804 |
| Random forest | 0.605824818 | 0.613512148 | 0.588746819 | 0.600505518 | 0.636140551 | 0.636140551 | 0.556253222 | 0.702670526 | 0.61791741 | 0.588859368 | 0.529801514 | 0.588966234 |